\documentclass[fleqn,usenatbib]{mnras}

\usepackage{newtxtext,newtxmath}
\usepackage[T1]{fontenc}

\usepackage[normalem]{ulem}

\DeclareRobustCommand{\VAN}[3]{#2}
\let\VANthebibliography\thebibliography
\def\thebibliography{\DeclareRobustCommand{\VAN}[3]{##3}\VANthebibliography}

\usepackage{graphicx}	% Including figure files
\usepackage{amsmath}	% Advanced maths commands
\usepackage{multirow}
\usepackage{svg}
\usepackage{tabularx}
\usepackage{comment}

\newcommand{\msun}{\mathcal{M}_{\sun}}
\newcommand{\orcid}[1]{\href{https://orcid.org/#1}{\includesvg[width=10pt]{orcid}}}

\title[Galactic SFH from \textit{Gaia} GCNS WDLF]{A new method to retrieve the star formation history from white dwarf luminosity functions -- an application to the \textit{Gaia} catalogue of nearby stars}

\author[M. C. Lam et al.]{
M. C. Lam$^{\orcid{0000-0002-9347-2298}}$\thanks{Contact e-mail: \href{mailto:mlam@roe.ac.uk}{mlam@roe.ac.uk}},
N. Rowell$^{\orcid{0000-0003-3809-1895}}$,
and H. W. Yeung$^{\orcid{0000-0002-4993-9014}}$
\\
Institute for Astronomy, University of Edinburgh, Royal Observatory, Blackford Hill, Edinburgh EH9 3HJ, UK
}

\date{Accepted 2025 December 16. Received 2025 December 15; in original form 2025 March 18}

\pubyear{2025}

\begin{document}
\label{firstpage}
\pagerange{\pageref{firstpage}--\pageref{lastpage}}
\maketitle

%%%%%%%%%%%%%%%%%%%%%%%%%%%%%%%%%%%%%%%%%%%%%%%%%%%%%%%%%%%%%%%%%%%%%%%%%%%%%%%%

\begin{abstract}
With the state-of-the-art \textit{Gaia} astrometry, the number of confirmed
white dwarfs has reached a few hundred thousand. We have reached the era where
small features in the white dwarf luminosity function~(WDLF) of the solar
neighbourhood can be resolved. We demonstrate how to apply Markov chain Monte
Carlo sampling on a set of pre-computed partial-WDLFs to derive the star
formation history of their progenitor stellar populations. We compare the
results against many published works using various types
of stars, including white dwarfs, main sequence stars, sub-giants and 
the entire stellar population. We find convincing agreements among most of the
methods for the broad features in the star formation history,
particularly at the intermediate age of 0.1-9\,Gyr.

\end{abstract}


%%%%%%%%%%%%%%%%%%%%%%%%%%%%%%%%%%%%%%%%%%%%%%%%%%%%%%%%%%%%%%%%%%%%%%%%%%%%%%%%

%%%%%%%%%%%%%%%%%%%%%%%%%%%%%%%%% BODY OF PAPER %%%%%%%%%%%%%%%%%%%%%%%%%%%%%%%%

\section{Introduction}
%%%%%%%%%%%%%%%%%%%%%%%%%%%%%%%%%%%%%%%%%%%%%%%%%%%%%%%%%%%%%%%%%%%%%%%%%%%%%%%%
White dwarfs~(WDs) are the final stage of stellar evolution of main
sequence~(MS) stars with zero-age MS~(ZAMS) mass less than $8\msun$. Since this
mass range encompasses the vast majority of stars in the Galaxy, these
degenerate remnants are the most common final product of stellar evolution.
Thus, they are a good population to study the history of star formation in the
Galaxy. At this late stage of stellar evolution, there is an inconsequential amount of nuclear
base burning to replenish the energy they radiate away~\citep{2010ApJ...717..183R}. As a consequence, the
luminosity and temperature decrease monotonically with time. The electron
degenerate nature means that a WD with a typical mass of $0.6\mathcal{M}_{\sun}$
has a similar size to the Earth, giving rise to their high densities, low
luminosities, and large surface gravities.

%%%%%%%%%%%%%%%%%%%%%%%%%%%%%%%%%%%%%%%%%%%%%%%%%%%%%%%%%%%%%%%%%%%%%%%%%%%%%%%%
The use of the white dwarf luminosity function~(WDLF) as a cosmochronometer was
first introduced by \citet{1959ApJ...129..243S}. Given the finite age of the
Galaxy, there is a minimum temperature below which no white dwarfs can reach in
a limited cooling time. This limit translates to an abrupt downturn in the WDLF
at faint magnitudes. Evidence of such behaviour was observed by
\citet{1979ApJ...233..226L}, however, they were not sure at the time whether it
was due to incompleteness in the observations~(e.g.,~\citealp{1984ApJ...282..615I}). A decade later,
\citet{1987ApJ...315L..77W} gathered concrete evidence for the downturn and
estimated the age\footnote{``Age'' refers to the total time since the oldest
WD progenitor arrived at the zero-age main sequence.} of the disc to be
$9.3 \pm 2.0$\,Gyr~(see also \citealt{1988ApJ...332..891L}). While most studies
focused on the Galactic discs~\citep{1989LNP...328...15L, 1992ApJ...386..539W,
1995LNP...443...24O, 1998ApJ...497..294L, 1999MNRAS.306..736K,
2012ApJS..199...29G, 2021A&A...649A...6G}, some worked with the stellar
halo~\citep{2006AJ....131..571H, 2011MNRAS.417...93R, 2017AJ....153...10M,
2019MNRAS.482..715L, 2021MNRAS.502.1753T}.
 
%%%%%%%%%%%%%%%%%%%%%%%%%%%%%%%%%%%%%%%%%%%%%%%%%%%%%%%%%%%%%%%%%%%%%%%%%%%%%%%%
As most WDs have a similar broadband colour to main sequence stars, they cannot be
identified using photometry alone. They are found from UV-excess, large
proper motion and/or parallax. Because of the strongly peaked surface gravity
distribution of WDs, photometric fitting for their intrinsic properties
is usable by assuming a surface gravity. WDs fitted in such a way are useful
statistically provided that the sample is not strongly selection biased. This
is demonstrated in various pre-\textit{Gaia}-era studies without accurate parallaxes
comparing photometric and spectroscopic solutions to calibrate the atmosphere
model~\citep{2019ApJ...871..169G, 2019ApJ...882..106G}. However,
this is no longer a preferred way since accurate \textit{Gaia} parallaxes allow
much more accurate determination of WD properties~\citep[e.g.][hereafter, GF21]{2021MNRAS.508.3877G}. Though it was still used in the WD catalogues of the
$Gaia$ Catalogue of Nearby Stars~(GCNS, see below).

The \textit{Gaia} satellite provides parallactic measurements for over a billion point
sources~\citep{2021A&A...649A...1G, 2021AJ....161..147B} of which $359,000$
are high confidence WD candidates~(GF21).
The availability of parallaxes allows much more accurate fitting, which is
particularly important when the surface gravity is unknown for the photometric
sample. This has completely revolutionized the field of WD sciences \citep{2024NewAR..9901705T}. 

%%%%%%%%%%%%%%%%%%%%%%%%%%%%%%%%%%%%%%%%%%%%%%%%%%%%%%%%%%%%%%%%%%%%%%%%%%%%%%%%
The Early Data Release 3~(EDR3) of \textit{Gaia} relies on 34 months of observations, it
represents an improvement on all fronts over DR2, with parallax measurements
being now on average 20-30 per cent more accurate and proper motion measurements
twice as accurate as in the previous
DRs~\citep{2021A&A...649A...1G, 2021A&A...649A...2L}. The $Gaia$ Catalogue of
Nearby Stars~(GCNS) contains 331\,312 objects within 100\,pc of the Sun, of
which 21\,848 are white dwarfs~\citep[][
see Section~\ref{sec:gcns} for detailed description
]{2021A&A...649A...6G}.

%%%%%%%%%%%%%%%%%%%%%%%%%%%%%%%%%%%%%%%%%%%%%%%%%%%%%%%%%%%%%%%%%%%%%%%%%%%%%%%%
This article is organized as follows: in Section~2, we go through the background
of WDLFs; in Section~3 we explain how WDs can be used to retrieve the SFH of a
population and introduce a new concept -- the partial WDLF. We explain the fitting
procedure in Section~4 and then apply it to the \textit{Gaia} data in Section~5. In
Section~6 we compare the results against previous works. We conclude and discuss
potential future work in Section~7.

\section{White Dwarf Luminosity Function}
%%%%%%%%%%%%%%%%%%%%%%%%%%%%%%%%%%%%%%%%%%%%%%%%%%%%%%%%%%%%%%%%%%%%%%%%%%%%%%%%
The WDLF is a common tool for deriving the age of a stellar population. The WDLF is
the number density of WD as a function of luminosity, and it is an evolving
function with time. Its shape and normalisation are determined from only a few
parameters. \citet{1987ApJ...315L..77W} compared an observed WDLF derived from
the Luyten Half-Second~(LHS) catalogue with a theoretical WDLF to obtain an
estimate of the age of the Galaxy for the first time with this technique.
\citet{1990ApJ...352..605N} examined WDLFs with various SFH scenarios, finding 
that the WDLF is a sensitive probe of the star formation history~(SFH) as
it shows signatures of star formation episodes such as bursts and lulls.
\citet{2013MNRAS.434.1549R} took it further to address this inverse problem
mathematically and showed some success in recovering the SFH of the solar
neighbourhood when compared against SFH computed from other methods. By
decomposing the disks and halo components of the Milky Way, we can have an
independent view of the star formation history revealed by only the WD
populations, where they are most useful in deriving the SFH of old stellar
populations~\citep{2011MNRAS.417...93R, 2017ASPC..509...25L}. Throughout this
work, we use lower case italicized $m$ for apparent magnitude, upper case
italicized $M$ for absolute magnitude, cursive upper case $\mathcal{M}$ for
mass, subscript $i$ and $f$ are reserved for \textit{initial} (progenitor) and
\textit{final} (WD). Hence, the summation indices are over $j$ throughout this
work. $t$ is for time and $\tau$ is the inverse cooling rate of WDs.

%%%%%%%%%%%%%%%%%%%%%%%%%%%%%%%%%%%%%%%%%%%%%%%%%%%%%%%%%%%%%%%%%%%%%%%%%%%%%%%%
The physical picture of getting a population of isolated WDs is
straightforward: the progenitor stars formed in their birth clusters
following a distribution of mass~($\mathcal{M}_i$) described by the initial
mass function~(IMF, $\phi$). Then, they spend their MS lifetime carrying out
nuclear burning~($t_{\mathrm{MS}}$), and the time they spend depends mainly
on their mass. Towards the end stage of the MS stellar evolution, stars lose
most of the atmosphere, modelled by the initial-final mass relation~(IFMR,
$\zeta$). Once they have become WDs, all that is left is to know how long it
has been cooling~($t_{\mathrm{cool}}$) in order to reach the current
luminosity~($M_\mathrm{bol}$). Most of the computations to account for these
physical processes are derived from pre-computed models. Particular care is
needed to interpolate and integrate over the grids of WD evolutionary models,
because they are both susceptible to significant rounding errors given the huge
dynamic ranges the variables cover. For example, in the case of a simple
starburst of $\mathcal{O}(10^6)$\,yrs, it requires a relative error tolerance
of $\texttt{epsabs} = \texttt{epsrel} = 10^{-10}$ for the \textsc{scipy} 
interpolator, a large number of breakpoints in the bounded integration interval
($\texttt{n\_points} = 10000$) and an arbitrarily large value of the upper bound
on the number of sub-intervals for integration ($\texttt{limit} = 10^{6}$) in
order to integrate properly for an old population with a short 
starburst. See Appendix~\ref{appexdix:integration-precision} for a more 
detailed description.

%%%%%%%%%%%%%%%%%%%%%%%%%%%%%%%%%%%%%%%%%%%%%%%%%%%%%%%%%%%%%%%%%%%%%%%%%%%%%%%%
When the luminosity function is \textit{properly} smoothed and weighted, and the
uncertainties accurately propagated, the parameterization using luminosity or
magnitude should give identical results to within the errors coming from
the interpolation over the model grid covering a large dynamic range of values.
In this work, we parametrize the computation with the bolometric magnitude so
the integral for a WDLF is written as

\begin{equation} \label{eqn:wdlf}
    n(M_{\mathrm{bol}}) = \int_{\mathcal{M}_l}^{\mathcal{M}_u}
        \tau(M_\mathrm{bol}, \mathcal{M}_f)
        \psi(t_0, M_\mathrm{bol}, \mathcal{M}_i, \mathcal{M}_f, Z)
        \phi(\mathcal{M}_i) d\mathcal{M}_i
\end{equation}
where $n$ is the number density, $\tau$ is the inverse cooling rate, $\psi$ is
the relative star formation rate, $\phi$ is the initial mass function; and their
dependent variables: $M_\mathrm{bol}$ is the absolute bolometric
magnitude, $\mathcal{M}_f$ is the WD mass, $t_0$ is the look-back time, $\mathcal{M}_i$ is
the progenitor MS mass, $Z$ is the metallicity, $\mathcal{M}_l$ is the minimum
progenitor MS mass that could have evolved in isolation into a WD in the given time,
and $\mathcal{M}_u$ is the maximum progenitor MS mass. The
normalisation of the IMF is set by integrating the IMF with the limits of $0.6$
and $8.0$\,${M}_\odot$. Hence, the SFH from this work is representative of
the stars with masses within that range.

%%%%%%%%%%%%%%%%%%%%%%%%%%%%%%%%%%%%%%%%%%%%%%%%%%%%%%%%%%%%%%%%%%%%%%%%%%%%%%%%
The inverse cooling rate
\begin{equation}
    \tau(M_\mathrm{bol}, \mathcal{M}_f) = \dfrac{dt_{\mathrm{cool}}}{dM_\mathrm{bol}} \left( M_\mathrm{bol}, \mathcal{M}_f \right)
\end{equation}
is a quantity taken from the pre-computed grid of cooling models. This rate is
also dependent on the internal structure and the chemistry of the core as well
as the atmosphere of the WDs. However, interpolation over these dependencies is
not possible with the available models, so their effects are not included in
this work, and is not included in the equation.

The relative star formation rate is expressed as a function of look-back time,
\begin{align}
\begin{split}
    &\psi(t_0, M_\mathrm{bol}, \mathcal{M}_i, \mathcal{M}_f, Z) =\\
    &\qquad\psi\left[t_0 - t_{\mathrm{cool}}\left(M_\mathrm{bol}, \mathcal{M}_f\right) - t_{\mathrm{MS}}\left(\mathcal{M}_i, Z\right)\right]
\end{split}
\end{align}
where $\psi$ is set to zero when $t_0$ exceeds the total time spent on the
main sequence evolution and the WD cooling time~(i.e.\ the onset of star 
formation). The absolute normalization is not needed when the total stellar
mass is coming from observations, given the relative normalization from the 
integrations are retained. The theoretical WDLF only requires a constant
multiplier (the total number density) to account for the normalization.

%%%%%%%%%%%%%%%%%%%%%%%%%%%%%%%%%%%%%%%%%%%%%%%%%%%%%%%%%%%%%%%%%%%%%%%%%%%%%%%%
The IFMR takes a simple form of
\begin{equation}
    \mathcal{M}_f = \zeta(\mathcal{M}_i),
\end{equation}
although there is evidence that more metal-rich stars lose more
envelope~\citep{2007ApJ...671..761K}, there is insufficient empirical data
to derive an IFMR at metallicity much lower or higher than solar abundance.

%%%%%%%%%%%%%%%%%%%%%%%%%%%%%%%%%%%%%%%%%%%%%%%%%%%%%%%%%%%%%%%%%%%%%%%%%%%%%%%%
All the calculations in this work adopt the following models: the IMF
is that of \citet{2003PASP..115..763C}, MS lifetime and metallicity are those from
PARSEC's solar metallicity model~\citep[Z=0.017, Y=0.279;][]{2012MNRAS.427..127B};
the IFMR is from \citet{2008MNRAS.387.1693C} and the WD cooling models and their
synthetic photometries with pure hydrogen atmosphere (DA) are from the Montreal
group's August 2020 version~\citep{2020ApJ...901...93B}\footnote{\url{http://www.astro.umontreal.ca/~bergeron/CoolingModels}}.
All the theoretical WDLFs and post-hoc bolometric magnitude uncertainties are
computed using \textsc{WDPhotTools}\footnote{\url{https://github.com/cylammarco/WDPhotTools}}~\citep{marco_c_lam_2022_6595029, 
2022RASTI...1...81L}.

%%%%%%%%%%%%%%%%%%%%%%%%%%%%%%%%%%%%%%%%%%%%%%%%%%%%%%%%%%%%%%%%%%%%%%%%%%%%%%%%
\section{Methods to Retrieve Star Formation History from a WD population}
\citet{1990ApJ...352..605N} were the pioneers in studying how the shape of
a WDLF is more sensitive to the time-dependency of the SFR than to the changes
in the IMF. They have attributed the broad bump at
$M_{\mathrm{bol}} \approx 10\,\mathrm{mag}$ to a recent burst of star formation
occurring at a look back time of 0.3\,Gyr. After a long hiatus, with new
high-quality photometric data available from, most importantly, 
SDSS~\citep{2000AJ....120.1579Y}, SuperCOSMOS~\citep{2001MNRAS.326.1279H} and
Pan-STARRS~1~\citep{2016arXiv161205560C}, there was increasing evidence that
such a feature exists, but it remained inconclusive due to the size of the
uncertainty in the  measurements~\citep{2006AJ....131..571H,
2011MNRAS.417...93R, 2019MNRAS.482..715L, 2024MNRAS.535.3611Q}.

%%%%%%%%%%%%%%%%%%%%%%%%%%%%%%%%%%%%%%%%%%%%%%%%%%%%%%%%%%%%%%%%%%%%%%%%%%%%%%%%
\subsection{Direct binning of age estimated from individual stars}
\citet[][hereafter, T14]{2014ApJ...791...92T} reported the SFH using a
spectroscopic sample of WDs complete to 20\,pc. They found two peaks of star
formation at around a look-back time of $\approx4.5$\,Gyr and at 
$\approx8.5$\,Gyr. \citet{2019ApJ...878L..11I} uses accurate parallactic 
measurements in \textit{Gaia} to study massive white dwarfs in the solar
neighbourhood as reported in \citet{2019Natur.565..202T} to study the effect of
crystallization. They found two peaks in the SFH from these massive WDs within
100\,pc, one at $\approx2.5$\,Gyr and one at
$\approx7$\,Gyr. This approach relies on the estimation of the age of each
individual star, so it is only applicable to a relatively small sample with
highly accurate parallactic measurements because of the mass-radius degeneracy
in the solution.

%%%%%%%%%%%%%%%%%%%%%%%%%%%%%%%%%%%%%%%%%%%%%%%%%%%%%%%%%%%%%%%%%%%%%%%%%%%%%%%%
\subsection{Inversion of the WDLF}
With a mathematical inversion method, it is possible to retrieve the SFH of a
stellar population. \citet{2013MNRAS.434.1549R} adopts the Richardson-Lucy 
deconvolution method~\citep{1972JOSA...62...55R, 1974AJ.....79..745L} that is 
typically used in image reconstruction. In this case, the images are the
two-dimensional histograms in the progenitor mass -- formation time plane and
the WD mass -- luminosity plane. However, it is prone to the amplification of
noise due to the existence of degenerate solutions that can yield the same
WDLF to within the uncertainty. Some forms of explicit regularization are
necessary to ensure smoothness in the solution. For example, 
\citet{2013MNRAS.434.1549R} defines a convergence
criterion~(section~\textsection 2.1.4) as a $1\%$ change in the
best fit $\chi^2$, beyond which they consider the improvement as overfitting.
They have  recovered a SFH in good agreement with other works, including the
peak at 0.3\,Gyr, and an enhanced star formation peaking at around 2\,Gyr, a
lull at 6\,Gyr and a broad strong star formation showing the combined thin and
thick disks between 7 and 10\,Gyr.

%%%%%%%%%%%%%%%%%%%%%%%%%%%%%%%%%%%%%%%%%%%%%%%%%%%%%%%%%%%%%%%%%%%%%%%%%%%%%%%%
\subsection{Forward modelling with parametric SFH}
\citet{2019ApJ...887..148F} investigated the SFH using
CFIS~\citep{2017ApJ...848..128I} and Pan-STARRS\,1 3$\uppi$
survey~\citep{2016arXiv161205560C}. CFIS covers $10,000$\,deg$^2$ and
$5000$\,deg$^2$ of the northern sky in u and r-band photometry to a depth of
$24.2$ and $24.85$ (AB\,mag) respectively. They fit the sample simultaneously
with three skewed Gaussian distributions that represent the thin disk, thick
disk and stellar halo population. They identified a peak SFH at $\approx8$\,Gyr
which is dominated by their thick disk population.

\citet[][hereafter, C23]{2023MNRAS.522.1643C} performs simulations
to evolve the stellar systems from zero age main sequence and dynamically evolve
them to derive their WDLFs. They found a constant star formation history from a
lookback time of 0 to 10\,Gyr. \citet[][hereafter, R25]{2025MNRAS.538.2548R}
extended the work to study the systematic biases comprehensively and concluded that
sharp features in the SFH derived from WDLFs are not statistically significant,
but there is a strongest period of star formation at around 3\,Gyr lookback time.

%%%%%%%%%%%%%%%%%%%%%%%%%%%%%%%%%%%%%%%%%%%%%%%%%%%%%%%%%%%%%%%%%%%%%%%%%%%%%%%%
\subsection{Forward modelling with non-parametric SFH~(this work)}
Another option to work on a statistical sample is to match a theoretical
WDLF based on an initial guess input SFH. However, a direct implementation to derive a
non-parametric SFH is extremely computationally heavy. In view of this, we
follow the mathematical construction of full spectrum fitting of galactic
spectra, and the use of partial CMD to derive the SFH of a stellar
population~\citep{2006A&A...459..783C}, to speed up the modelling significantly.
In both of these methods, they derived various physical properties, e.g.\ SFH
and metallicity, of the stellar population. In the case of WDLF inversion in
this work, SFH is the only independent variable (function). Full spectrum
fitting works by comparing the observed spectrum against a combination of basis
model spectra of a set of pre-computed simple stellar populations. Because of
the large possible number of degenerate solutions and the sensitivity to noise,
regularization is necessary to avoid erroneously amplified
solutions. For example, \texttt{pPXF}~\citep{2023MNRAS.526.3273C} implements an 
explicit regularisation parameter, which can be interpreted as a convergence 
criterion as in \citet{2013MNRAS.434.1549R}. Alternatively, a Markov chain
Monte-Carlo method can be used to mitigate ``overfitting'' (e.g.\ in 
\texttt{Prospector},~\citealp{2021ApJS..254...22J}). This automatically comes
with the statistics of the solutions. This approach requires immense
computational power and, thus, it has not been explored in previous work.

%%%%%%%%%%%%%%%%%%%%%%%%%%%%%%%%%%%%%%%%%%%%%%%%%%%%%%%%%%%%%%%%%%%%%%%%%%%%%%%%
In order to apply the fitting method to compute the SFH from WDLFs, we introduce
the partial WDLF~(pWDLF). It is inspired by the partial CMD designed
in~\citet{2006A&A...459..783C}, we use these pWDLFs as basis models for fitting.
Similar to their method, the word \textit{partial} refers to a view of the
system over a small time range. In the context of this work, a pWDLF corresponds
to a WDLF from a stellar population with an intense star formation at the given
time and with no star formation before or after that. The pre-computation of a
set of pWDLFs introduces a heavy overhead to the analysis. However, this can
significantly reduce the computation demand during the fitting stage. The
pre-computation can also allow much simpler reanalysis or fitting multiple
WDLFs since the set of pWDLFs is invariant.

\citet[][hereafter, I08]{2008ApJ...682L.109I} noted that at
$M_{\mathrm{bol}} < 13$, the shape of a WDLF is almost independent of the star
formation rate. R11 disagreed on that conclusion; from our set of pWDLFs~(see
Section~\ref{sec:application}), it shows that I08's statement only holds
if an observer can only see a perfectly smooth WDLFs in the range of 
$9 < M_{\mathrm{bol}} < 13$, and without the knowledge of the completeness level
of the observation. A short period of star formation corresponds to a
prominent peak in a WDLF, changes in the SFR in $\sim$$0.25-2$\,Gyr~(corresponding
to the magnitude range above) would appear as a bump in the WDLF.

%%%%%%%%%%%%%%%%%%%%%%%%%%%%%%%%%%%%%%%%%%%%%%%%%%%%%%%%%%%%%%%%%%%%%%%%%%%%%%%%
\section{SFH from fitting pWDLFs}
\label{sec:fitting}
The fitting for the SFH is done in two steps: the first one uses a Markov-Chain
Monte Carlo method, and the second step refines the solution with a least-squares
method. The free parameters to be fitted are the weights of the pWDLFs required 
to compose a theoretical WDLF that matches the observed WDLF. The weight of
each pWDLF in reconstructing the WDLF is directly proportional to the SFH
over the corresponding time range.
\begin{comment}
% NR: removed; not clear what it means
\sout{provided that the set of pWDLFs was computed using the same normalising factor
from the IMF and IFMR.}
\end{comment}

This can be written as

\begin{equation}
    n(M_\mathrm{bol}) = \sum_j w_j \times n_j(t, M_\mathrm{bol})    
\end{equation}

where the $w_j$ is the weight of the $j$-th pWDLF ($n_j$) with a short burst of
star formation at time $t$.

In this work, we fit with 58 pWDLFs~(see Section~\ref{sec:application}). 
The total number of steps was set to 1\,000\,000 with a burn-in of
10\%. The 31.73 and 68.27 percentiles are computed as the upper and lower 1-sigma 
uncertainties of the solution. The solution is then fed into the 
\texttt{scipy.optimize.least\_squares} minimisation function to compare against
the observed one with the \texttt{ftol}, \texttt{xtol} and \texttt{gtol} set to
\texttt{1E-10}. The refinement only leads to little change to the solution, but
this guarantees the accuracy and repeatability of the solution.

%%%%%%%%%%%%%%%%%%%%%%%%%%%%%%%%%%%%%%%%%%%%%%%%%%%%%%%%%%%%%%%%%%%%%%%%%%%%%%%%
The likelihood function to be maximized is essentially minimizing the $\chi^2$
between the observed and the reconstructed WDLFs weighted by the variance from
the observed WDLF:
\begin{equation}
    \chi^2 = \sum \frac{\left[n(M_\mathrm{bol}) - n_\mathrm{obs}\right]^2}{\sigma_n^2}
\end{equation}
where $n(M_\mathrm{bol})$ is the reconstructed WDLF, $n_\mathrm{obs}$ is the
observed WDLF, $\sigma_n$ is the standard deviation in $n_\mathrm{obs}$, and the
summation is over magnitude.

%%%%%%%%%%%%%%%%%%%%%%%%%%%%%%%%%%%%%%%%%%%%%%%%%%%%%%%%%%%%%%%%%%%%%%%%%%%%%%%%
\subsection{Test case - noiseless}

We demonstrate the performance of the use of pWDLF in two scenarios:
(i)~exponentially increasing star formation rates at five truncation ages, and
(ii)~two bursts where the broad and weak burst is fixed in time while the strong
narrow bursts are superposed at five different ages, as well as one with all
these profiles combined. The bin size in both time (number of pWDLFs) and
magnitude (number of data points in a WDLF) are the set used in analyzing the
empirical \textit{Gaia} EDR3 data~(see Section~\ref{sec:magnitude_bin_size} for how
the binning is determined).

%%%%%%%%%%%%%%%%%%%%%%%%%%%%%%%%%%%%%%%%%%%%%%%%%%%%%%%%%%%%%%%%%%%%%%%%%%%%%%%%
\subsubsection*{Exponentially increasing SFR with a truncation}
\begin{figure}
  \includegraphics[width=\columnwidth]{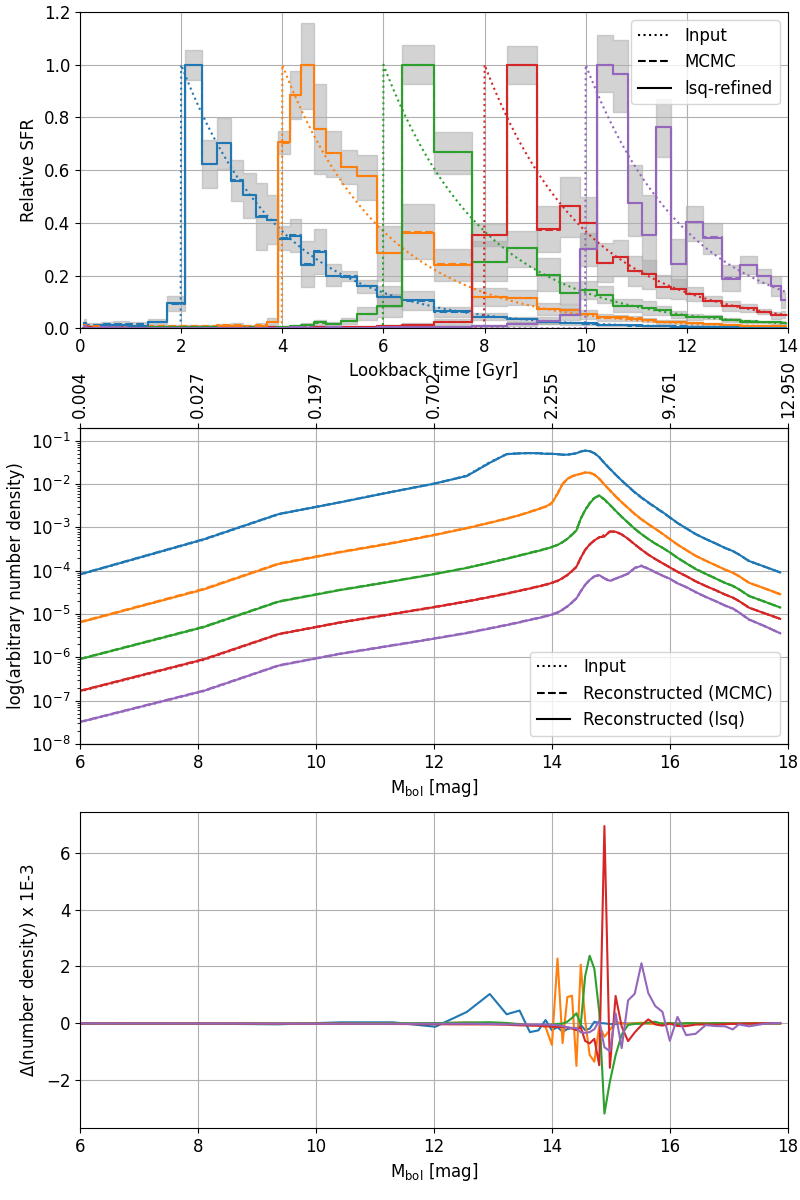}
  \caption{In the top two panels, the input are plotted with dotted line; the
  marginalized results from the MCMC are plotted with dashed line; and the
  refined solutions with a least-square minimization method using a set of
  perturbed MCMC results as initial conditions are plotted in solid lines.
  The least-squares ~(lsq) method is to validate that the solutions are not
  trapped inside a local minimum. The MCMC and lsq solutions are almost
  identical. Top: the recovered SFH and reconstructed SFH of a set of mock
  stellar populations with exponentially increasing star formation rates. Each
  of the population has an exponential constant of 3\,Gyr, the SFHs were
  truncated at a look-back time of 2, 4, 6, 8, and 10\,Gyr. See 
  Section~\ref{sec:magnitude_bin_size} for the explanation for the varying bin
  size in the lookback time. Middle: the input and reconstructed WDLFs. Bottom: The difference
  in the number density between the lsq solution and the input.}
  \label{fig:exponential_sfh}
\end{figure}

The set of unrealistic SFH as shown in Fig.~\ref{fig:exponential_sfh} tests the
capability of the MCMC method in recovering slowly varying trends as well as
rapidly truncated features in the SFH. The five input SFHs have an exponentially
increasing SFH peaked at a look back time of $2$, $4$, $6$, $8$ and $10$\,Gyr,
where the exponential constant is $3$\,Gyr. We can see excellent agreement in
the recovered SFH to the input SFHs. Except for the bins adjacent to
the truncations of the SFH, all but one bin in the purple curve have
discrepancies within two standard deviations from the input values. 

%%%%%%%%%%%%%%%%%%%%%%%%%%%%%%%%%%%%%%%%%%%%%%%%%%%%%%%%%%%%%%%%%%%%%%%%%%%%%%%%
\subsubsection*{Multiple bursts}
\begin{figure}
  \includegraphics[width=\columnwidth]{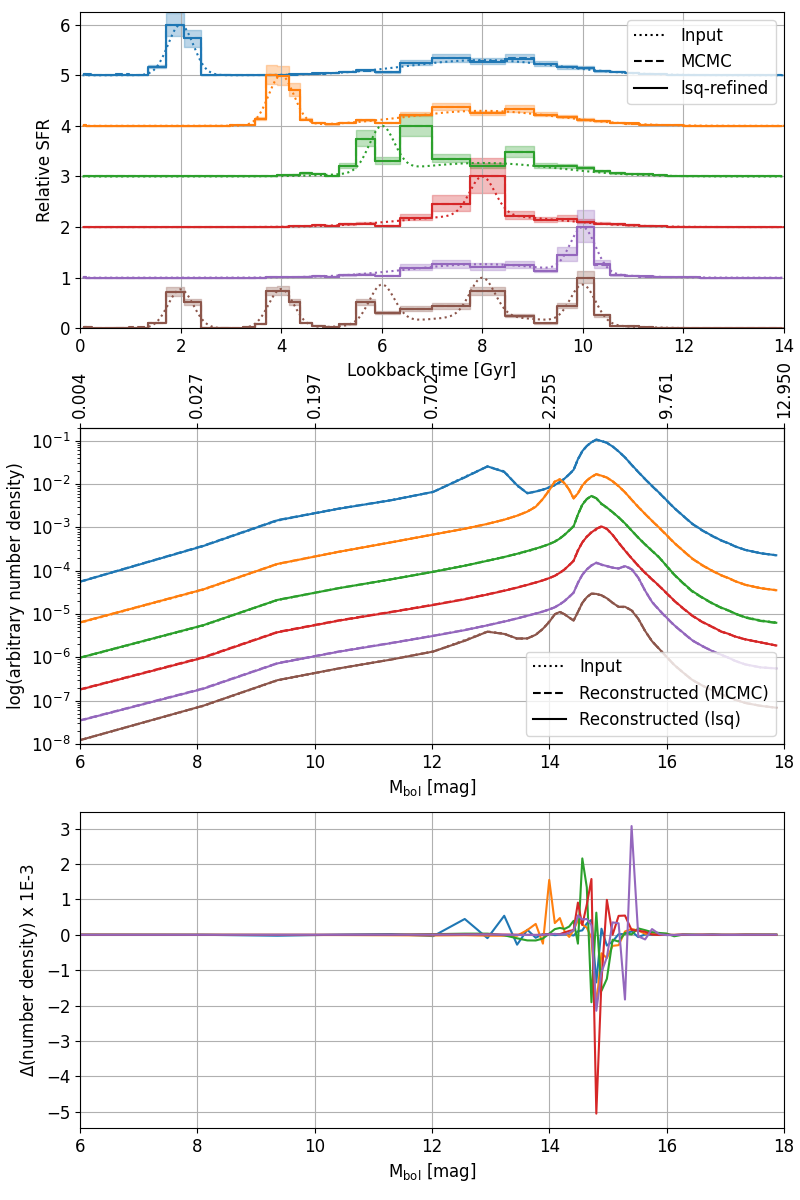} 
  \caption{The recovered SFH and reconstructed SFH of a set of mock stellar
  populations with a dual-burst SFH (blue, orange, green, red and purple), and
  the last one (brown) is a superposition of five short and one broad SFH.
  See Fig.~\ref{fig:exponential_sfh} for the descriptions of the three sets
  of legends.}
  \label{fig:bursts_sfh}
\end{figure}

The set of WDLFs shown in Fig.~\ref{fig:bursts_sfh} demonstrates that the method
can retrieve both sharp and
broad features whether they are distinctly separated or directly on top of each
other. The last set (brown) comprises all five sharp peaks and the broad
feature. It shows that some correlated results are not completely
accounted for at around $6-7$\,Gyr (M$_\mathrm{bol}\approx14-15$\,mag). This is
coming from the fact that the pWDLFs are very similar as the WDLFs evolve the
slowest at these magnitudes, as is obvious from the similarity in the green and
red curves in Fig.~\ref{fig:bursts_sfh}. We plan to address this issue in future
work by considering the colour information~(see the discussion in 
Section~\ref{sec:conclusion}) which should relieve some degeneracy issues in the
solutions.

%%%%%%%%%%%%%%%%%%%%%%%%%%%%%%%%%%%%%%%%%%%%%%%%%%%%%%%%%%%%%%%%%%%%%%%%%%%%%%%%
\section{Application to the Early \textit{Gaia} Data Release 3}
\label{sec:application}
Upon deriving an SFH from a WDLF, one of the most important unsolved problems
is ``How much information can we obtain?''. We address in this work how, as a
first step, to achieve the maximal sampling in the magnitude space to extract
as much information from a WDLF as possible while avoiding the amplification of
noise. Any attempt to retrieve a signal at finer intervals than the
resolution of the data is essentially amplifying random noise. In this section,
we define the optimal bin sizes in bolometric magnitude, and how this naturally
defines the lookback time bin sizes.

\subsection{WDs from the GCNS}
\label{sec:gcns}
%%%%%%%%%%%%%%%%%%%%%%%%%%%%%%%%%%%%%%%%%%%%%%%%%%%%%%%%%%%%%%%%%%%%%%%%%%%%%%%%
%
% Comments on the completeness / contamination of the GCNS sample
%

The GCNS contains 21\,848 objects within 100\,pc that pass the survey selection
criteria and are assigned a probability of being a WD ($P_{\text{WD}}$) higher
than 0.5. Classification is based on a random forest method applied to the
$Gaia$ proper motions,  parallaxes and integrated photometry ($G$,
$G_{\text{BP}}$, $G_{\text{RP}}$), and trained on several different catalogues
of known WDs \citep[see][section~5.8]{2021A&A...649A...6G}.
The nominal $Gaia$ G band magnitude limit of 20.7 implies that the 100\,pc
sample is incomplete for WDs with absolute G$\gtrsim$15.7, which includes part
of the local WD population slightly beyond the peak in the WDLF. This is
accounted for in their WDLF by the use of the 1/V$_{\mathrm{max}}$ method.
The magnitude completeness arising from the complex survey
selection function is accounted for with the supplementary data available with
GCNS\footnote{\textsc{maglim.dat.gz} from \url{https://cdsarc.u-strasbg.fr/viz-bin/cat?J/A+A/649/A6}}. 
This selection is not applied to the main catalogue, but it is applied to the
WD sample when deriving their WDLFs~(Hambly, 2025, priv. comm.). Hence, some
of the 21\,848 sources are not included in the WDLF catalogue. This reduces
the sample size to contain only 19\,113 unique sources. Furthermore, the GCNS
WDLF catalogue contains all objects it classified as WDs that have more than
1\% chance of lying within 100\,pc~(see the end of this subsection). Hence,
most of the objects at a relatively large distance in the catalogue only have
a fractional contribution to the statistics. Their weighted sum gives us an
effective number of 16\,063.29 objects.
Of the 19\,113 GCNS WDLF WDs, 18\,796 are matched to a GF21 WD within
1~arcsec with parallax larger than 8\,mas, of which 15\,909 have parallax
larger than 10\,mas.

While the GCNS 100\,pc sample is certainly not as well understood as the
40\,pc sample \citep{2024MNRAS.527.8687O}, it is nevertheless a useful
catalogue for demonstrating our SFH recovery method.

The WDs are fitted with the pure hydrogen model from the Montreal
group~\citep{2019ApJ...876...67B} ignoring the effects of varying
the H/He atmospheric composition and surface gravity. The GCNS work
interpolated the model grid to look up $G$-band bolometric corrections as a
function of ($G - G_{\mathrm{RP}}$) to map M$_\mathrm{G}$ to M$_{\mathrm{bol}}$,
at a fixed surface gravity of $\log(g)=8.0$. Because of the different colour
evolution as a function of the chemistry of the atmosphere, opting to fit a
pure hydrogen atmosphere limits the accuracy of the estimation of the cooling
ages as mixed hydrogen-helium atmosphere model give a
shorter cooling age than the pure hydrogen model~\citep{2022ApJ...934...36B}.
\begin{comment}
Overall, this treatment can lead to an incorrect WD radius
by as much as a factor of 2, hence, an incorrect luminosity by a factor
$\sim$$4$, causing spurious features in the WDLF.
\end{comment}

When computing the WDLF, the GCNS adopted a fixed scaleheight of 365\,pc, a
value found from the GCNS itself. This choice can lead to significant effect
on the volumetric correction based on the 1/V$_{\mathrm{max}}$ method for
deriving the WDLF.
%
% NR updated text:
Specifically, the scaleheight (or equivalently, vertical velocity dispersion) increases
with total stellar age \citep{2022A&A...658A..22R}, which manifests as an increased
average scaleheight for faint WDs \citep{2019MNRAS.484.3544R}, resulting in an
underestimate of their spatial density in the local volume if a fixed value is assumed.
% While \citet{2016MNRAS.461.2100T} found this to have a significant effect on the
% estimation of the WD mass distribution, the impact on the WDLF has not been studied in detail.
Previous works on empirical
WDLFs for the disc have generally adopted a fixed scaleheight, typically of 
250pc~\citep{2006AJ....131..571H,2011MNRAS.417...93R, 2019MNRAS.482..715L}.
An age-dependent scaleheight has been adopted in C23 and R25, although
the sensitivity to different values has not been studied in detail (and see 
Section~\ref{ssec:otherMethodsWds} for discussion of some issues in their
scaleheight).

This sample also assumed all WDs have come from isolated stellar evolution.
However, recent works suggest that as much as 40\% of single
WDs have come from mergers \citep{2020A&A...636A..31T, 2020ApJ...898...84K}.
They are products
of stars that have gone through stellar interactions in the progenitor
systems~\citep[e.g.\ ][]{2013A&ARv..21...59I, 2023arXiv231117145H}. We do not
attempt to address this effect in this work.

The data available from the GCNS WDLF is a catalogue of WDs where instead of
having each row corresponding to one object, each row contributes
as $1\%$ of a WD to the
WDLF. When performing photometric fitting of the WDs, the posterior distribution
of the distance of each WD is stored for each percentile. Each row represents
a $1\%$ chance of the WD at that distance and bolometric magnitude. The catalogue
reports the central $99$ percentiles for each WD. The maximum volumes are computed
for each percentile for every WD. In some cases, the tail of the distribution
exceeds 100\,pc~(the distance limit of GCNS), these WDs only contribute a small
fraction of a WD to the total WDLF. The density normalization and the
uncertainties in the WDLF have to be rescaled properly when summing for the
total WDLF because directly summing the solution from each
row will lead to an amplification of $\sim$$100$ times in the density.
Similarly, the error bars would be $100$ times too small if uncorrected. Since
the catalogue has retained enough information on the bolometric magnitude and 
distance, we can choose any bin size to present the GCNS WDLF as required to
optimally extract the SFH. However, one downside of the catalogue is that it
does not come with the uncertainties of the derived bolometric magnitude, so
we refit the photometry that has taken into account the photometric and
parallactic uncertainties, intrinsic distribution in the surface gravity and
the accuracy in using synthetic photometry~(see below). In this work, we use
the GCNS $M_\mathrm{bol}$ and the refitted uncertainties.

%%%%%%%%%%%%%%%%%%%%%%%%%%%%%%%%%%%%%%%%%%%%%%%%%%%%%%%%%%%%%%%%%%%%%%%%%%%%%%%%
\subsection{Bin size}
\label{sec:magnitude_bin_size}
At the Nyquist sampling rate~\citep{1949IEEEP..37...10S},
there should be two sampling points in the space of one full-width at half
maximum~(FWHM). Assuming the noise is Gaussian, one FWHM is equivalent to
$2.355$ standard deviations ($\sigma$). Thus, the optimal sampling rate is
$1.1775\sigma$ at the given magnitude.
The maximal information within a WDLF that can be extracted
is thus limited by the level of uncertainties in the luminosity determination,
in this work it is the bolometric magnitude. The uncertainty in the
bolometric magnitudes from photometric fitting acts as the smoothing kernel
that degrades the true signal. By constructing a one-to-one relation between
the bolometric magnitude and its uncertainty, we can determine the resolution
with which to retrieve true information from the WDLF.

%%%%%%%%%%%%%%%%%%%%%%%%%%%%%%%%%%%%%%%%%%%%%%%%%%%%%%%%%%%%%%%%%%%%%%%%%%%%%%%%
\subsubsection{Bolometric magnitude}
We have identified two main contributions to the uncertainties in the
bolometric magnitudes, with which we require to derive the size of the
smoothing kernel as a function of the bolometric magnitude.

\paragraph{Intrinsic distribution of surface gravity \hfill\\}
\label{sec:logg_intrinsic_dispersion}
Given the simple treatment adopted in the GCNS sample selection, a representative
``intrinsic'' distribution in the surface gravity should include non-DA WDs, we
take the 40\,pc WDs from Table 1 of \citet{2024MNRAS.527.8687O} to get an average
$\log(g)=8.0607$ with a dispersion of $0.2783$.

\paragraph{Uncertainties in photometric fitting \hfill\\}
\label{sec:logg_photometric_fitting}
From the lower panel of Fig.~14 of \citet{2021MNRAS.508.3877G}, the dispersion
in the difference between spectroscopic surface gravities and the photometric ones
is about 0.03~dex based on the width of the envelope.

\paragraph*{Total uncertainties and dispersion \hfill\\}
In order to obtain the estimates of the uncertainties in the fitted bolometric
magnitudes, we used \textsc{WDPhotTools} to fit the GCNS WD samples in the
three \textit{Gaia} filters at seven surface gravities. These seven values range from
$\log(g) - 3\sigma_{\log(g)}$ to $\log(g) + 3\sigma_{\log(g)}$
in an increment of one $\sigma_{\log(g)}$. While the $\sigma_{\log(g)}$ is the
sum in quadrature of the intrinsic distribution of surface gravity and the
dispersion in surface gravity coming from photometric fitting~
(Section~\ref{sec:logg_intrinsic_dispersion} and \ref{sec:logg_photometric_fitting}).

From the seven fitted
bolometric magnitudes, we can approximate the dispersion with:
\begin{multline}
  \sigma_{M_{\mathrm{bol}}} = \frac{1}{6} \times \sum_{-3 \leq j \leq 3, j \neq 0} \Biggl\{ \frac{M_{\mathrm{bol}}(\log(g)=7.998 + i\sigma_{\log(g)})}{|j|} \\
  - M_{\mathrm{bol}}(\log(g)=7.998) \Biggr\}
\end{multline}

The Gaussian smoothing kernel is described by the sum of the fitting
uncertainties, the model uncertainties and the distribution coming from the
assumption of fixed surface gravity in quadrature~(blue scattered points in top
panel of Fig.~\ref{fig:magnitude_resolution}). The weighted average uncertainty
as a function of the bolometric magnitude is computed using the inverse maximum
volume as weights as it is used to weight each data point to the WDLF\footnote{
the bin centres of the bolometric magnitude used to compute the average are
1.5,  2.5,  3.5,  4.5,  5.25,  5.75,  6.25,  6.625, 6.85 to 17.25 in increment
of 0.2 and the last bin is centred at 17.475}.

%%%%%%%%%%%%%%%%%%%%%%%%%%%%%%%%%%%%%%%%%%%%%%%%%%%%%%%%%%%%%%%%%%%%%%%%%%%%%%%%
The use of WDLF relies on the assumption that WD shares similar properties
at the same luminosity. Based on such an underlying assumption, we measure
the mean of these dispersions weighted by the $1/\mathrm{V}_{\mathrm{max}}$
as provided in the GCNS WD catalogue. Then they are fitted with a spline as
a function of bolometric magnitudes~(see Fig.~\ref{fig:magnitude_resolution}
and Table~\ref{tab:magnitude_resolution} in the appendix for the interpolation
presented in the middle panel of Fig.~\ref{fig:magnitude_resolution}) to get
the average uncertainty as a function of the bolometric magnitude.

%%%%%%%%%%%%%%%%%%%%%%%%%%%%%%%%%%%%%%%%%%%%%%%%%%%%%%%%%%%%%%%%%%%%%%%%%%%%%%%%
\subsubsection{Lookback time}
\begin{figure}
    \includegraphics[width=\columnwidth]{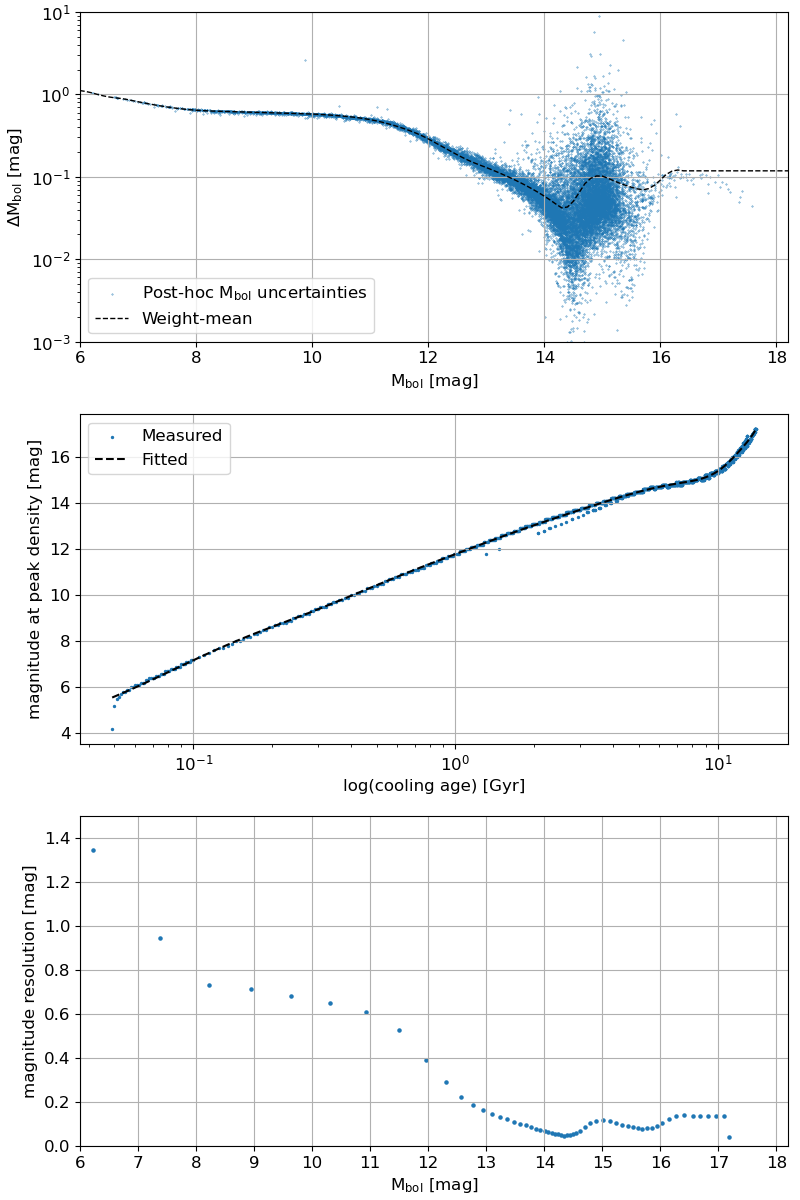}
    \caption{Top: the total uncertainties in the bolometric magnitude as a 
    function of bolometric magnitude~(blue). Due to the small
    number of data points, the weighted average at magnitudes fainter than
    $16.5$\,mag are not computed. At these magnitudes, the uncertainties at the
    last bin centred at $16.45$\,mag is used. Middle: the bolometric magnitude
    at which the peaks of the pWDLF are located. Bottom: the bolometric
    magnitude resolution of the WDLF computed using the fitting uncertainties
    from the top figure and the added uncertainties coming from the synthetic
    photometry and the assumption of fixed surface gravity~(see text). The
    low resolution in the bright end comes from a combination of large
    uncertainties in the observed magnitudes, the lack of UV photometry and the
    sensitivity of the solution to the surface gravity.}
    \label{fig:magnitude_resolution}
\end{figure}

\begin{figure}
    \includegraphics[width=\columnwidth]{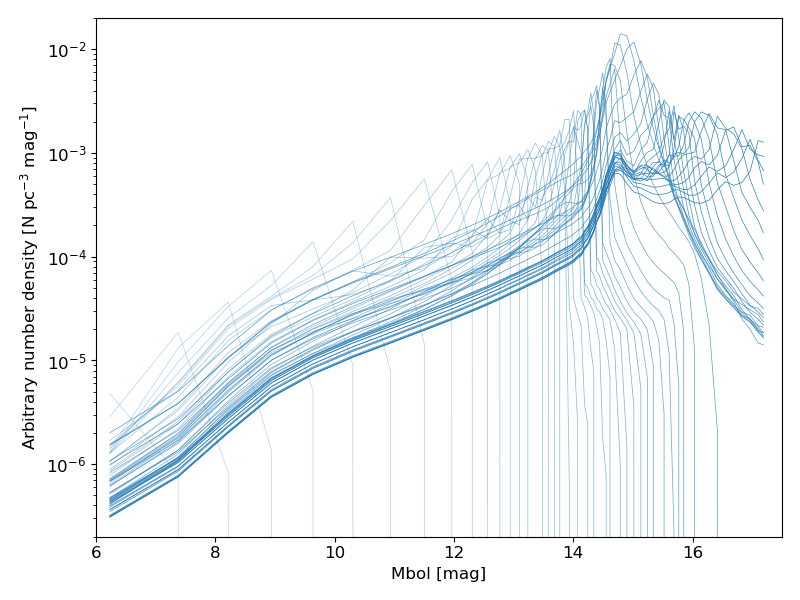}
    \caption{The set of partial WDLF (basis functions) that the peaks are
    separated by the magnitude resolution as found in
    Fig~\ref{fig:magnitude_resolution}. At older ages, double peaks start to
    emerge. However, there is always a dominant peak that is at least half an
    order of magnitude higher in the contribution. All pWDLFs have the same
    relative normalization coming directly from the integrations using various
    input model.}
    \label{fig:basis_pwdlf}
\end{figure}

The WDLFs with short bursts of star formation are strongly peaked at a single
magnitude (see, for example, the set of pWDLFs in Fig.~\ref{fig:basis_pwdlf}).
Thus, together with the monotonically cooling nature of WDs,
there is a one-to-one mapping of the cooling age of the population and the peak
magnitude (see middle panel of Fig.~\ref{fig:magnitude_resolution}). Although
at older ages, with the peak magnitude beyond $15$\,mag, a WDLF starts to show
double peaks, there is always a dominant peak that is at least half an order of
magnitude higher in the number density. Thus, we can reliably relate the two
relations: peak magnitude--age and peak magnitude--magnitude-resolution. Since
the peak magnitude has a one-to-one mapping to the age of the system, we can
obtain the time resolution in which we construct the set of the basis pWDLFs
that would allow the extraction of the SFH at the maximum sampling
rate~(i.e.\ the minimum bin size) without the amplification of noise.

%%%%%%%%%%%%%%%%%%%%%%%%%%%%%%%%%%%%%%%%%%%%%%%%%%%%%%%%%%%%%%%%%%%%%%%%%%%%%%%%

%%%%%%%%%%%%%%%%%%%%%%%%%%%%%%%%%%%%%%%%%%%%%%%%%%%%%%%%%%%%%%%%%%%%%%%%%%%%%%%%

%%%%%%%%%%%%%%%%%%%%%%%%%%%%%%%%%%%%%%%%%%%%%%%%%%%%%%%%%%%%%%%%%%%%%%%%%%%%%%%%
\subsection{Star Formation History}
\label{sec:sfh}
\begin{figure}
    \includegraphics[width=
    \columnwidth]{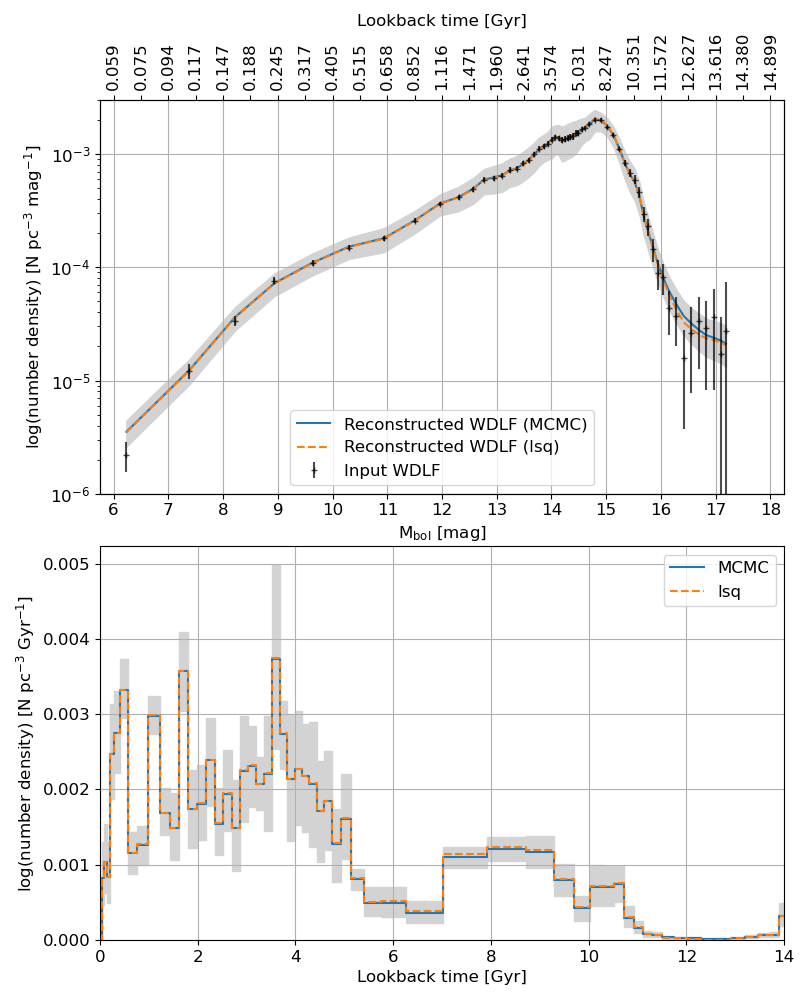}
    \caption{Top: the reconstructed WDLF using MCMC method on the pWDLFs is
    shown in blue, where the set of solutions is being further refined with a
    least-squares method is shown in orange. The x-axis on the top side shows only the
    cooling age WDs but not the total age. A luminosity function has
    marginalized over the progenitor masses, so the total age of WDs is not
    linearly mapped to the axis of this figure. Bottom: The weights of the
    pWDLFs transformed to the unit of per billion year by dividing by the bin
    widths, and by normalizing with the number density to get the unit of per
    cubic parsec. The raw weights are found from MCMC sampling and finishing
    with a least-squares minimization.}
    \label{fig:sfh_optimal}
\end{figure}

By applying the set of basis functions to the fitting method as described in
section \textsection\ref{sec:fitting}, because the set of pWDLFs is
normalized the same way, the relative contribution translates to the relative
SFH of the solar neighbourhood. The absolute normalization can be found by
matching the reconstructed WDLF to the integrated number density of the input
WDLF. We have found multiple phases of
enhanced star formation using this novel method, these bursts and lulls are all
in good agreement with previous works found using other methods and stellar 
populations~(see Fig.~\ref{fig:sfh_optimal} and the next section).
We can see a broad feature of star formation at $0-5$\,Gyr,
and another one at $7-11$\,Gyr. There are some prominent bursty star
formations, see a more in-depth analysis in the next subsection to assess the
presence of these peaks.

%%%%%%%%%%%%%%%%%%%%%%%%%%%%%%%%%%%%%%%%%%%%%%%%%%%%%%%%%%%%%%%%%%%%%%%%%%%%%%%%
%The 20\,pc subset\ shows the contribution of the WDs within the 20\,pc distance,
%it should \textit{not} be interpreted as a 20\,pc sample where the maximum
%volume (upper integral limit) for each WD would be different to that of a
%100\,pc sample. We do not compute the maximum volume in this work, hence we have
%opted for the word ``subset''. We show this subset for a more
%meaningful comparison of other works that probed much smaller volumes~(see 
%Section~\ref{sec:comparison}). As already pointed out in the previous section,
%we caution readers to pay particular attention in drawing conclusions from any 
%information older than $\sim$$9$\,Gyr due to, most likely, some correlated
%signals unaccounted for.

The tabulated input and best-fit WDLFs from this work can be found in
Appendix~\ref{appexdix:reconstructed-wdlf}.

%%%%%%%%%%%%%%%%%%%%%%%%%%%%%%%%%%%%%%%%%%%%%%%%%%%%%%%%%%%%%%%%%%%%%%%%%%%%%%%%
\subsection{Bootstrapping}
In order to assess whether the features in the computed SFH are significant, we
perform a bootstrapping exercise following closely the recipe described in
Section 4 of R25. We resample five input models of the construction of the
pWDLFs: (1)~The exponent of the IMF ($2.3$) is resampled with a standard
deviation of $0.1$~(e.g.\ in~\citet{2018ApJ...860L..17E, 2025MNRAS.538.2548R}.
(2)~The MS lifetime is multiplied by a factor of $(1+p)$ where $p$ is drawn
from a Gaussian distribution with a standard deviation of
$0.048$~\citep{2000MNRAS.315..543H}. (3 \& 4)~The IFMR is a linear equation where
the gradient has an uncertainty of $0.004$ and the constant has an uncertainty
of $0.011$~\citep{2008MNRAS.387.1693C}. (5) The WD cooling time is multiplied
by a factor of $(1+p)$ where $p$ is drawn from a Gaussian distribution with a
standard deviation of $0.06$~\citep{2023MNRAS.522.1643C, 2024MNRAS.527.3602C}.

\begin{figure}
    \centering
    \includegraphics[width=\columnwidth]{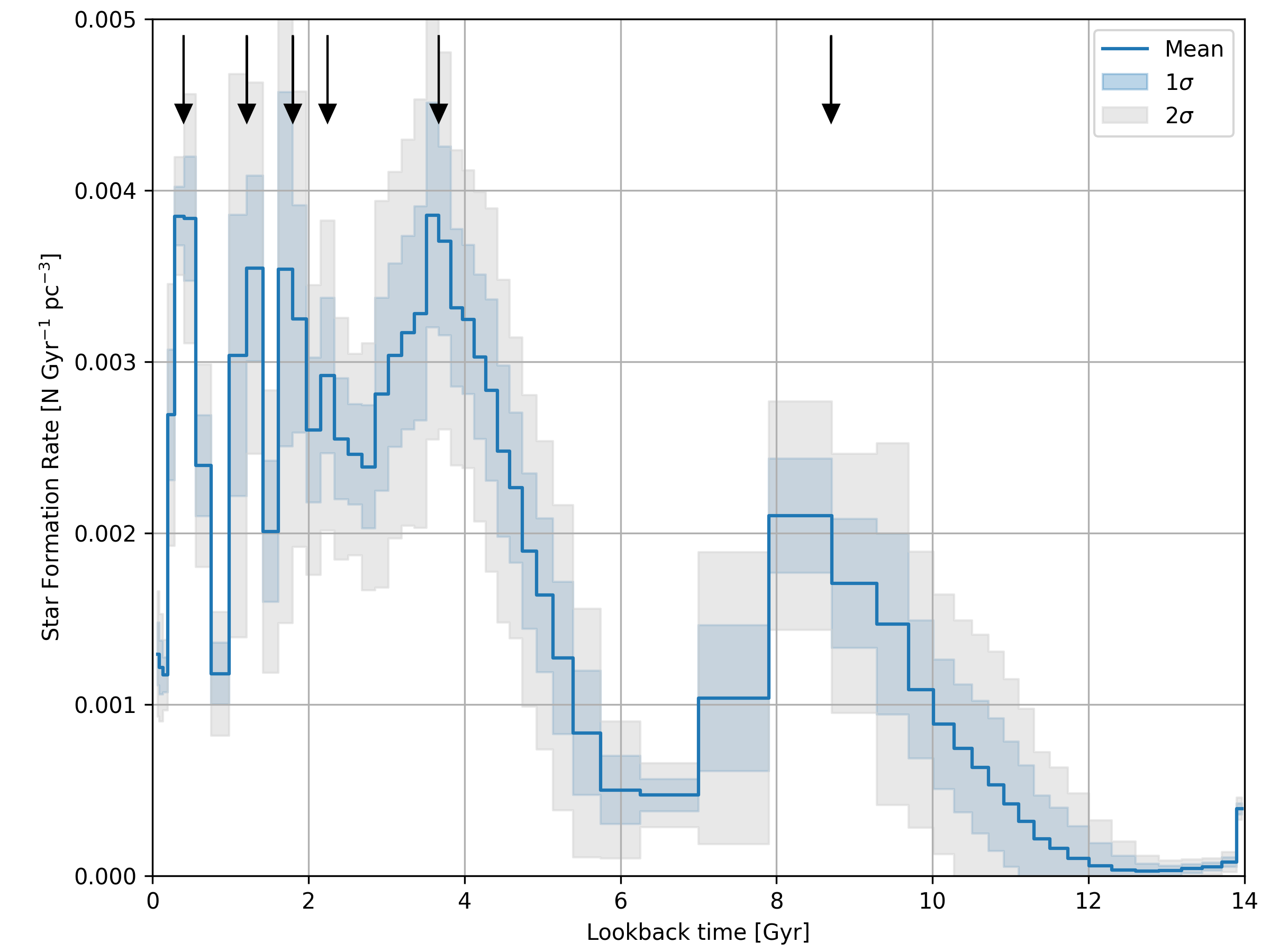}
    \caption{
    The mean SFR from the 1000 bootstrap solutions as a
    function of lookback time. It shows very similar shape and feature to the
    best-fit solution, with many of the noisy features removed. The blue and
    grey bands show the $1\sigma$ and $2\sigma$ distribution of the bootstrap
    solutions. The six arrows mark the peaks of enhanced star formation at
    around $0.40$, $1.21$, $1.80$, $2.25$, $3.67$ and $8.7$\,Gyr. See more in
    Figure~\ref{fig:full_sample_peaks}.
    }
    \label{fig:bootstrap_mean}
\end{figure}

\begin{figure}
    \centering
    \includegraphics[width=\columnwidth]{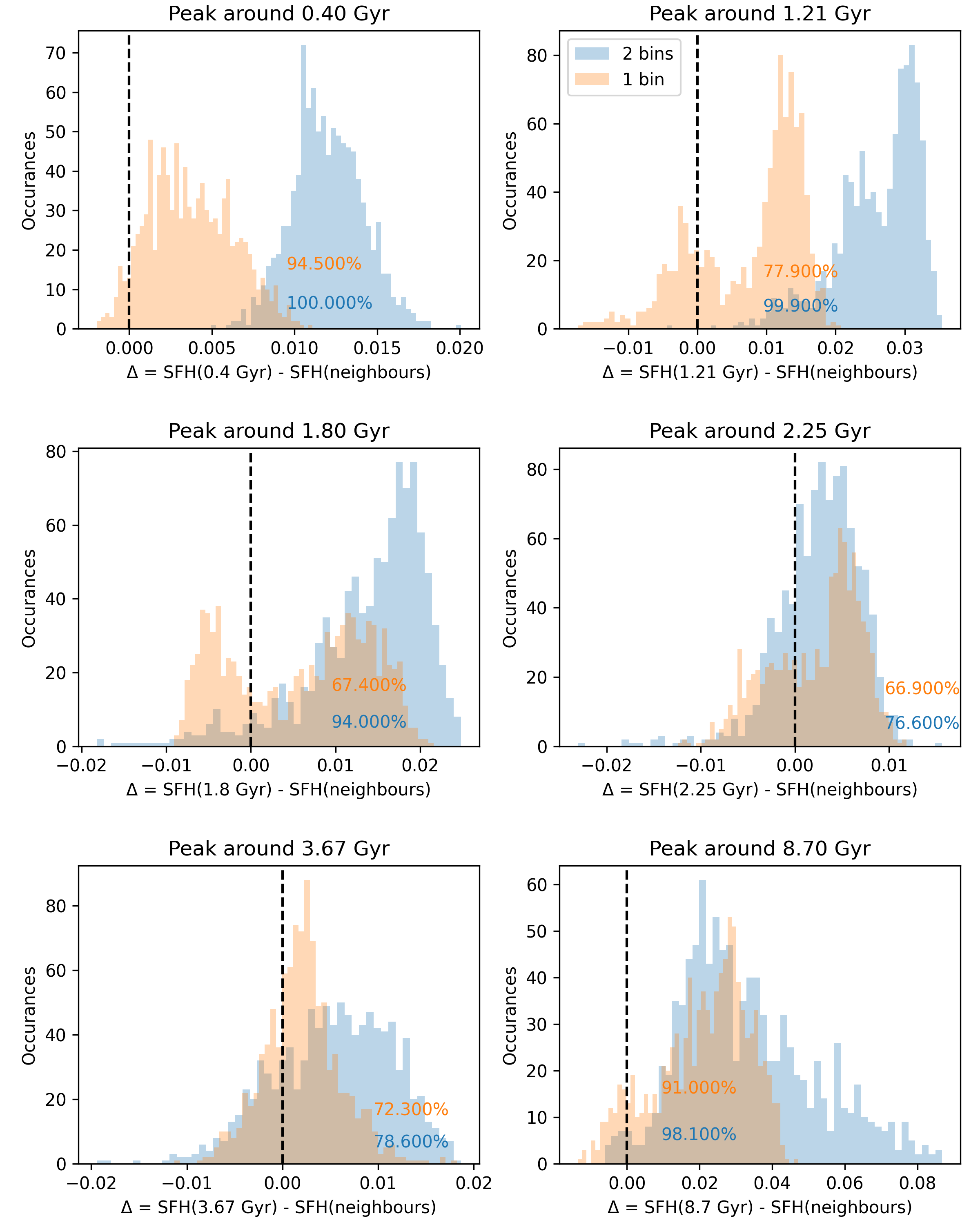}
    \caption{
    Each peak labelled in Figure~\ref{fig:bootstrap_mean} is analysed in a
    1-bin and 2-bin test. The peaks at $0.40$, $1.21$, $1.80$ and $8.7$\,Gyr have
    more stars formed per unit volume than their neighbours in over $90\%$ the
    bootstrapped solutions. Both the peaks and the neighbours use the time
    elapsed by the peaks to compute the number density in the unit of
    N\,pc$^{-3}$. The 1-bin tests show lower chance of these peaks appearing,
    this is due to the peaks drifting back and forth in the time-axis in the
    bootstrapping process when the MS lifetime and WD cooling time are varied.
    Hence, the relatively broad features appear in only the 2-bin test but not
    the 1-bin. The peaks around $2.25$ and $3.67$\,Gyr only appear in
    $\sim$$75\%$ of the bootstrapped solutions, so they only have a slightly
    better than random chance to be present.
    }
    \label{fig:full_sample_peaks}
\end{figure}

%%%%%%%%%%%%%%%%%%%%%%%%%%%%%%%%%%%%%%%%%%%%%%%%%%%%%%%%%%%%%%%%%%%%%%%%%%%%%%%%
We construct $1000$ sets of pWDLFs by drawing from those five distributions.
These sets are computed in the same time and bolometric magnitude resolution
as in the set used in Section~\ref{sec:sfh}. Each of these was fitted with
the same procedure as done previously, except for shorter chains of length
100000 and burn-in of 10000 steps, since we already have a good starting point
by using the best-fit solution. The arithmetic means and standard deviations
from the bootstrapping solutions are plotted in Figure~\ref{fig:bootstrap_mean}.
It naturally presents smoother solutions than the best-fit solution where some
of the peaks no longer appear in the bootstrapped average solution. In this
analysis, we refer the narrow features of 1 to 2 bins in width as peaks.
Whereas, the broader features are clearly visible after the smoothing and
broadening of information in the time-axis by averaging all the bootstrapped
solutions. From the $1000$ solutions, we perform a basic analysis on the
persistence of these peaks. In Figure~\ref{fig:full_sample_peaks}, it shows
the distributions of the difference in the number density (N pc$^{-3}$)
between the peak and the neighbours. Six peaks from the full sample are
selected, as marked in Figure~\ref{fig:bootstrap_mean}. Given that all WDs
from the WDLF have to contribute to the total SFH, the bootstrapped solutions
are mostly about shifting the real peaks back and forth in time rather than
dampening the peaks. This makes the peaks broader in the averaged bootstrapped
solutions. Henceforth, each peak is defined as the most prominent bin or two
most prominent adjoining bins. The neighbours are the bins immediately before
and after the peaks. The difference, $\Delta$, is the difference in the number
densities from integrating over the time elapsed in the peaks and the average
neighbours' SFH integrated over the same time elapsed by the peaks. The 1-bin
and 2-bin cases are to cover both short bursts of star formation and the
slightly more gradual ones.

%%%%%%%%%%%%%%%%%%%%%%%%%%%%%%%%%%%%%%%%%%%%%%%%%%%%%%%%%%%%%%%%%%%%%%%%%%%%%%%%
There are two clear broad features, one is in the range of $0-6$\,Gyr, while the
other one is at $7-12$\, Gyr. For the peaks, we find the number of stars formed
at $0.40$, $1.21$, $1.80$ and $8.7$\,Gyr higher than their neighouring bins over
$90\%$ the time. The peaks at $0.40$, $1.21$ and $1.80$\,Gyr are showing much
more prominence in the 2-bin case than the 1-bin. This is due to the peaks
drifting back and forth in the time-axis in the bootstrapping process when the
MS lifetime and WD cooling time are varied. Hence, the relatively broad features
appear in only the 2-bin test but not the 1-bin. This is also obvious by eye
from Figure~\ref{fig:bootstrap_mean} where gaps are clearly visible between
those three peaks in the smoothed and broadened SFH. At $2.25$\,Gyr, there is
a slightly better than random chance of having a peak. At $3.67$\,Gyr, the peak
on top of the broad feature between three and $6$\,Gyr appears in around $78\%$
of the samples.

%%%%%%%%%%%%%%%%%%%%%%%%%%%%%%%%%%%%%%%%%%%%%%%%%%%%%%%%%%%%%%%%%%%%%%%%%%%%%%%%
\section{Comparison and discussion}
\label{sec:comparison}

\subsection{WDLF}
\label{sec:inversion}
%%%%%%%%%%%%%%%%%%%%%%%%%%%%%%%%%%%%%%%%%%%%%%%%%%%%%%%%%%%%%%%%%%%%%%%%%%%%%%%%
In order to validate the results in this paper, we have compared the
resulting SFH with the equivalent result obtained from an
independent method applied to the same input WDLF.

\begin{figure}
    \includegraphics[width=\columnwidth]{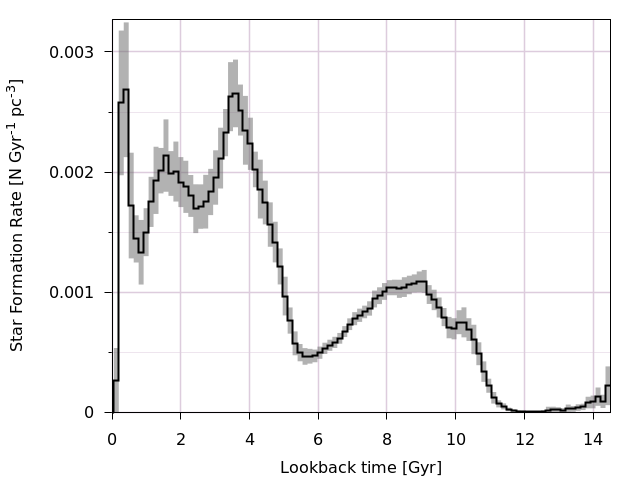}
    \caption{Star formation rate obtained by inversion of the GCNS WDLF using the
    algorithm presented in \citet{2013MNRAS.434.1549R} and described in 
    section~\ref{sec:inversion}.}
    \label{fig:nr_sfr}
\end{figure}

\begin{figure}
    \includegraphics[width=\columnwidth]{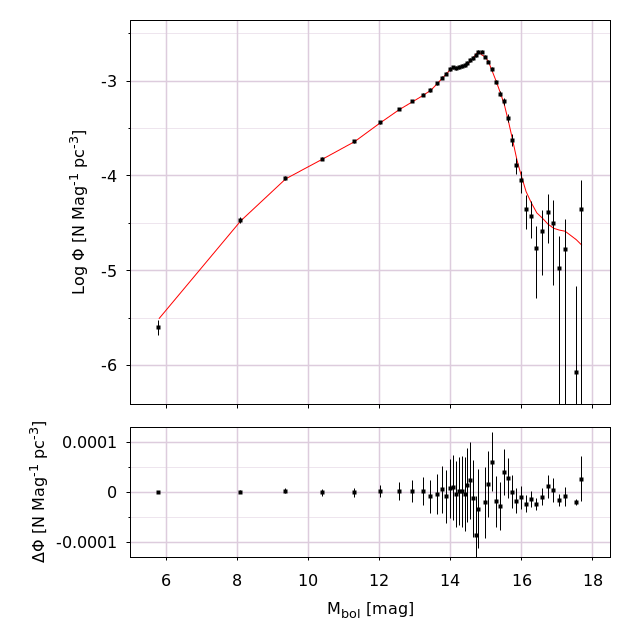}
    \caption{Comparison between the observed (black points) and model (red line)
    WDLF obtained from the inversion algorithm, and corresponding to the star
    formation history presented in figure~\ref{fig:nr_sfr}.}
    \label{fig:nr_wdlf}
\end{figure}

%%%%%%%%%%%%%%%%%%%%%%%%%%%%%%%%%%%%%%%%%%%%%%%%%%%%%%%%%%%%%%%%%%%%%%%%%%%%%%%%
\citet{2013MNRAS.434.1549R} presented a method to estimate the star formation
history from WDLFs using an inversion algorithm on the integral 
Equation~\ref{eqn:wdlf}. Their method is based on the expectation-maximization
algorithm, which is similar in principle to Richardson-Lucy deconvolution and is
used to obtain maximum-likelihood solutions to inverse problems in the presence 
of missing data. In the present application, this is the unknown distribution
of WD mass as a function of magnitude.
We applied their inversion algorithm to the GCNS dataset to compute the SFH using
the same astrophysical inputs and bolometric magnitude binning. For improved
characterisation of the statistical uncertainty we performed 200 inversions based
on 200 independent resamplings of the WDLF. The mean and sample standard deviation
of the resulting SFH is presented in Figure~\ref{fig:nr_sfr}, with the
corresponding reconstructed WDLF shown in Figure~\ref{fig:nr_wdlf}. This SFH
accounts for stars with masses in the range $0.6$--$7 M_{\sun}$.
The uncertainty is known to be underestimated, due to degeneracies around rapid
changes in the star formation rate. Systematic errors are also significant and arise
from the choice of the inputs, particularly the WD cooling models, as well as
unmodeled effects such as binarity and metallicity variations among the progenitor
population. This is explored in part in \citet{2013MNRAS.434.1549R}.

%%%%%%%%%%%%%%%%%%%%%%%%%%%%%%%%%%%%%%%%%%%%%%%%%%%%%%%%%%%%%%%%%%%%%%%%%%%%%%%%
The result from this independent inversion method shows
near-perfect agreement in the SFH over the entire cosmic timeline (see Figure~\ref{fig:comparison}). The four
main broad features are all recovered at the same lookback time. Given the
two recoveries use identical sets of MS lifetime model, WD cooling model, IMF,
IFMR, the remarkably similar SFHs are a strong indication that the recovered
signals are real.

\subsection{Other methods using WDs \label{ssec:otherMethodsWds}}
%%%%%%%%%%%%%%%%%%%%%%%%%%%%%%%%%%%%%%%%%%%%%%%%%%%%%%%%%%%%%%%%%%%%%%%%%%%%%%%%
%
Significant recent studies of the local SFH using WDs have
focused on the 40pc sample, which is now well characterised thanks to $Gaia$ data
\citep{2024MNRAS.527.8687O}.
C23 developed a novel approach to estimating the SFH based on
modeling the cumulative distribution of absolute G magnitude, thus avoiding the
binning of stars.
%
% binarity and merger remnants, cooling delays due to crystallisation
Their detailed analysis of systematics in the modeling inputs leads them to conclude
that the SFH within 40pc is consistent with uniform over the last
10.5 Gyr.
A similar conclusion is reached in the follow-up study of R25, which expanded the
analysis of C23 to consider three alternative methods to constrain the star formation
history using WDs.
Their direct age estimation avoids forward modeling using an assumed parametric
SFH, although requires temperatures and masses for individual stars
and thus is not suited to large photometric WD samples that lack spectra.
The resulting SFH (reproduced in Figure~\ref{fig:comparison}) has a maximum at 2--4 Gyr, although their solution is consistent with a uniform SFH within 2$\sigma$.
Both C23 and R25 adopt a scaleheight versus age relation in their simulations that
rises linearly before saturating at 140\,pc for all stars older than \textasciitilde7 Gyr.
This is based on a transformation of the observed vertical velocity dispersion versus age
relation for their 40pc WD sample, but is nevertheless significantly at odds with many 
other studies based both on legacy surveys
% Harris: 200-900pc
% Juric: (300pc/900pc).
% Kilic: 40pc WDs, 
(e.g.~\citealt{2006AJ....131..571H,2008ApJ...673..864J,2017ApJ...837..162K}) and the 
latest Gaia data
% 2019MNRAS.482.4570G (age-averaged WD scaleheight of 230pc)
% 2019MNRAS.485.5573T (WD has a scaleheight of 200pc: Fig.9)
% 2022MNRAS.511.3863E (260pc/693pc)
% 2023Galax..11...77V (\textasciitilde$280$pc/$800$pc)
(e.g.~\citealp{2019MNRAS.482.4570G,2019MNRAS.485.5573T,2022MNRAS.511.3863E,2023Galax..11...77V}),
which find age-averaged scaleheights around 200-300pc for the thin disc and
700-900pc for the thick disc.

%%%%%%%%%%%%%%%%%%%%%%%%%%%%%%%%%%%%%%%%%%%%%%%%%%%%%%%%%%%%%%%%%%%%%%%%%%%%%%%%
\citet{2019ApJ...878L..11I} performed an analysis by directly
computing the star formation history of a population of massive WDs selected
from \citet{2019Natur.565..202T}. The choice of using only massive WDs
removes the degeneracy issues arising from the choice of metallicity models
of the progenitors; and the fact that massive WDs are remnants of the most
massive stars that could have turned into WDs, where their progenitor lifetime
is short compared to the WD cooling age. They found an abrupt start of star
formation at $\sim$7\,Gyr and an enhanced star formation around 2-3\,Gyr. They
also note that they find a burst at $\sim$0.4\,Gyr when applying their method
on the 25\,pc sample from~\citet{2017ASPC..509...59O}. This recent burst was not
seen in all but one previous work -- R13's. Interestingly enough, we have
recovered the same peak with the GCNS sample using the R13 inverse
modelling method (Section~\ref{sec:inversion}) and the pWDLF forward modelling
method~(this work). However, this is only a by-eye comparison,
the relative strength and the location of the peaks in the SFH have to be
corrected for the larger merger fraction for higher mass stars~\citep{2020A&A...636A..31T, 2024ApJ...974...12J} as well as the large
cooling anomaly~\citep{2019ApJ...886..100C, 2024Natur.627..286B}.

%%%%%%%%%%%%%%%%%%%%%%%%%%%%%%%%%%%%%%%%%%%%%%%%%%%%%%%%%%%%%%%%%%%%%%%%%%%%%%%%
\citet{2021MNRAS.502.1753T} analysed a 100\,pc sample of WDs from
Torres et al.~(2019) with a completeness of $91\%$ at 20.5\,mag. This sample
contains 95 WDs that are disentangled into thin disk, thick disk and stellar
halo components. The data with their selection criteria applied is commonly
referred to as a high-velocity sample, but their treatment of the data
selection is much more sophisticated, making use of a supervised
machine learning method based on Random Forest techniques in eight-dimensional
space. They found a cut-off age of $12\pm0.5$\,Gyr where the peak of star
formation happened at $\sim$11\,Gyr. The star formation continued for about
4\,Gyr. $13\%$ of their WDs are younger than 7\,Gyr which is puzzling given
their kinematic choice, spurious WDs from the thin and thick disk should not
have contributed to such a fraction of contamination. They suggested that
individual analysis will be required to unravel the origin of these objects.

%%%%%%%%%%%%%%%%%%%%%%%%%%%%%%%%%%%%%%%%%%%%%%%%%%%%%%%%%%%%%%%%%%%%%%%%%%%%%%%%
\subsection{MS stars}
There are several works concerning the SFH of the solar neighbourhood. We
have selected the following three because their sample selections are similar
to that of the GCNS sample with little directional dependence.

%%%%%%%%%%%%%%%%%%%%%%%%%%%%%%%%%%%%%%%%%%%%%%%%%%%%%%%%%%%%%%%%%%%%%%%%%%%%%%%%
\citet{2006A&A...459..783C} used the \textit{Hipparcos} catalogue of stars
within 80\,pc of the Sun and brighter than $V=8$\,mag. The restrictive selection
has significantly excluded the effects due to photometric or kinematic
incompleteness. They computed the SFH by minimizing the differences between the
colour-magnitude diagram of the population from the Hipparcos data and the
sum of the partial CMD~(the concept that inspired this work to use partial
WDLFs). They recovered the strongest star formation at a lookback time of
$2-3$\,Gyr, and a slowly decreasing SFH up to a lookback time of $6$\,Gyr where
there is a sharp drop in SFR until a lookback time of $10-12$\,Gyr. Their SFH
shows excellent agreement to our work, except for the strong peak at the most
recent time~(see more in Section~\ref{sec:conclusion}).

%%%%%%%%%%%%%%%%%%%%%%%%%%%%%%%%%%%%%%%%%%%%%%%%%%%%%%%%%%%%%%%%%%%%%%%%%%%%%%%%
\citet{2007ApJ...665..767R} used the Hipparcos catalogue to calibrate
bias in the \citet{2005ApJS..159..141V} volume-limited spectroscopic
sample of 1\,039 FGK dwarfs in the solar neighbourhood with estimates of
their age and metallicity. They went through thorough selection criteria,
which resulted in two samples. Each sample has limiting absolute magnitudes
between $4$ and $\sim$$6$\,mag, with a distance limit of $\sim$$30$\,pc, and MS mass
in the range $\sim1.25$ to $\sim0.8$\,M$_{\sun}$. Their results are a direct
number count of stars in a volume-limited survey that only include late F to
early K stars, so their SFH as plotted in Fig.~\ref{fig:comparison} is a simple
normalisation by multiplying by an arbitrary value to allow comparison of the
\textit{shape}, particularly the peaks and troughs of the SFH. The most notable
peaks from their work are at $2$ \& $4$\,Gyr. Two narrow peaks at $6$ \&
$8$\,Gyr are also present.

%%%%%%%%%%%%%%%%%%%%%%%%%%%%%%%%%%%%%%%%%%%%%%%%%%%%%%%%%%%%%%%%%%%%%%%%%%%%%%%%
\citet{2018IAUS..330..148B} used the TGAS data to perform an initial
analysis with the technique of synthetic colour-magnitude diagram-fitting, they
found a recent star burst at a lookback time of $\sim$2--3\,Gyr, an enhanced
star formation $\sim$6--10\,Gyr. Figure~\ref{fig:comparison} shows their
\textit{mildly corrected} solutions.

%%%%%%%%%%%%%%%%%%%%%%%%%%%%%%%%%%%%%%%%%%%%%%%%%%%%%%%%%%%%%%%%%%%%%%%%%%%%%%%%
\citet{2019A&A...624L...1M} selected from \textit{Gaia} DR2 2\,890\,208 stars
with mean $G_{\mathrm{DR2}}<12$, estimated to be $97\%$ complete. Only stars
with proxy-absolute magnitude $G_{\pi}$ brighter than $10$\,mag are considered in
their second filtering in order to remove all brown dwarfs and white dwarfs in
their analysis. They applied a non-parametric method that uses an approximate
Bayesian computation~\citep{2017A&C....19...16J} algorithm to compute their
merit function by comparing the \textit{Gaia} data against the Besancon Galaxy model
fast approximation simulations~\citep{2018A&A...620A..79M}). They have found
that their analysis is most affected by the thick disk modelling and the
stellar evolution models. From their four choices of fitting configurations of
the \textit{Gaia} sample, they see a general trend of decreasing star formation from
$\sim$10\,Gyr to 6\,Gyr, followed by an enhanced star formation of 4\,Gyr in
duration with a peak at $\sim$2.5\,Gyr. They also found a sharp and fast drop
in the star formation rate in the most recent 1\,Gyr.

%%%%%%%%%%%%%%%%%%%%%%%%%%%%%%%%%%%%%%%%%%%%%%%%%%%%%%%%%%%%%%%%%%%%%%%%%%%%%%%%
\citet{2021MNRAS.501..302A} uses stars brighter than $G = 15$
within 100 pc of the Sun from Gaia DR2, using parallaxes and photometry to
reconstruct the age and metallicity distribution via a Bayesian hierarchical model. 
They infer a primary peak of star formation about 10 Gyr\,ago, then a local
minimum at around 8\,Gyr ago. It followed by a second peak of star formation
at $\sim$5\,Gyr, and thereafter a roughly constant star formation rate up to
the present time.

%%%%%%%%%%%%%%%%%%%%%%%%%%%%%%%%%%%%%%%%%%%%%%%%%%%%%%%%%%%%%%%%%%%%%%%%%%%%%%%%
\citet{2024A&A...687A.168G} applied a CMD-fitting procedure on
the GCNS sample to infer a dynamically evolved SFH, including effects of stellar
evolution and selection. They found that there are a few bulks of star formation.
The earliest one was at around $\sim$$10$\,Gyr ago, followed by another one
at $\sim$$6-8$\,Gyr. There is a local minimum at a lookback time of $\sim$$4$\,Gyr.
After that, there is a renewed bulk of star formation continuing to about half
a billion years ago.

%%%%%%%%%%%%%%%%%%%%%%%%%%%%%%%%%%%%%%%%%%%%%%%%%%%%%%%%%%%%%%%%%%%%%%%%%%%%%%%%
\citet{2025A&A...697A.128D} used a forward modelling approach
to infer jointly the disc’s SFH and IMF. We take the SFH from the thin-disk
solution of G13P-13 from Table 2 of their work for comparison. They found a
strong star formation at a lookback time of $9-10$\,Gyr, which slowly ceased to
a minimum at $6-7$\,Gyr. The SFR climbed back to a second period of enhanced
star formation with the maximum at $3-4$\,Gyr and reached a minimum at the most
recent time. It shows some similarity to \citet{2019A&A...624L...1M}, see the
comparison between them in \citet{2025A&A...697A.128D}.

\subsection{Subgiants}
%%%%%%%%%%%%%%%%%%%%%%%%%%%%%%%%%%%%%%%%%%%%%%%%%%%%%%%%%%%%%%%%%%%%%%%%%%%%%%%%
\citet{2022Natur.603..599X} used subgiants to map the metallicity, angular
momentum and SFH of the Milky Way. The distance they probe is much larger
than this work so it is expected that the recovered SFH shows different
features. Nevertheless, they observe one dominant peak at 10.5\,Gyr and two
small peaks at 2 and 4\,Gyr.

\begin{figure}
  \includegraphics[width=\columnwidth]{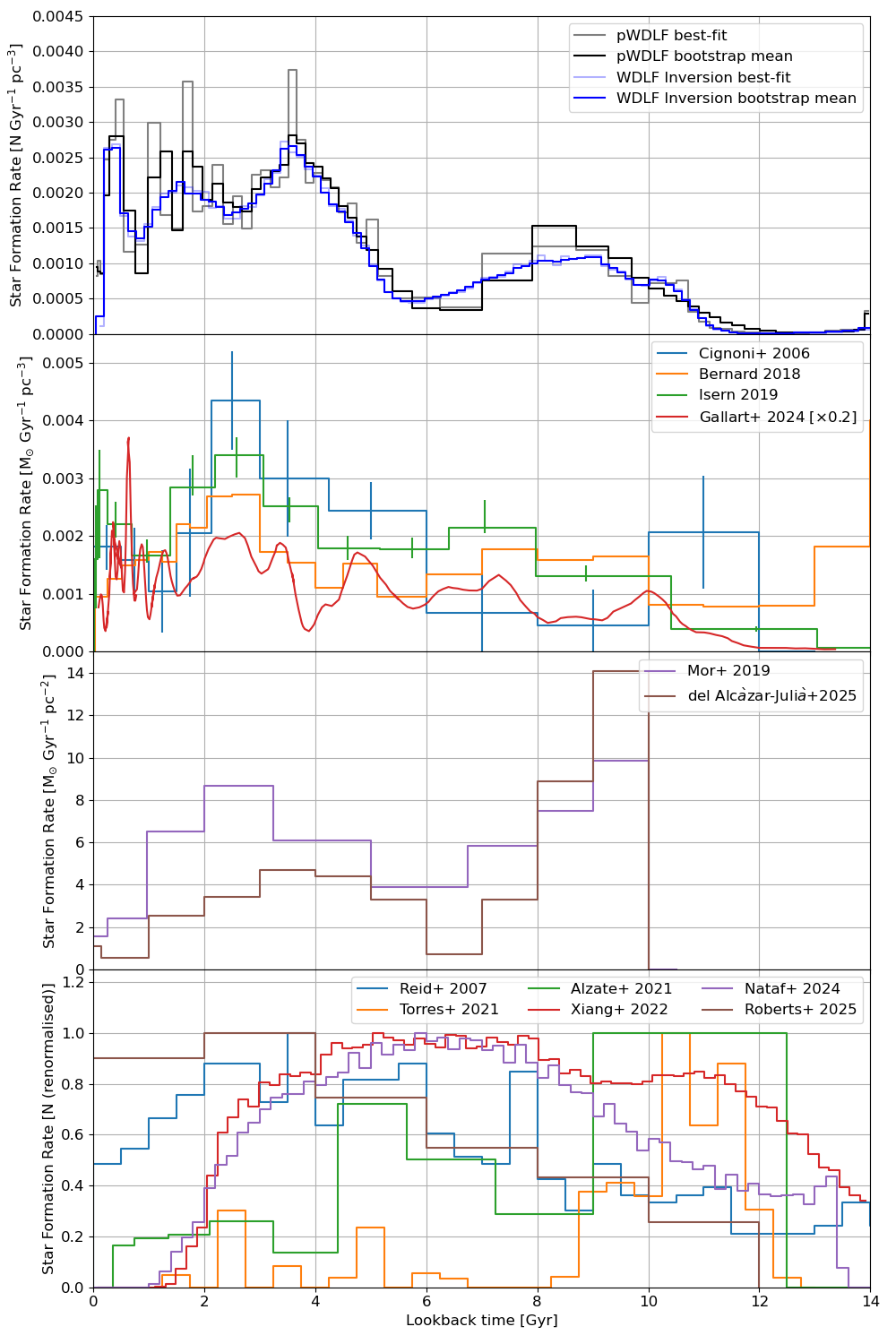}
  \caption{Comparison of the recovered SFH against previous works, see body
  text of Sec~\ref{sec:comparison} for details. The subplots are 
  grouped by their units in reporting the SFH, and the labelling is in
  chronological order. Top: SFH retrieved with the
  pWDLF method and the updated R13 method from this work. The SFH is reported
  in the unit of number of stars formed per unit time per unit volume.
  Second: The SFH using four independent methods and samples reported in the
  unit of solar mass formed per unit time per unit volume. Third: The SFH using
  two independent methods and samples reported in the unit of solar mass formed
  per unit time per unit projected area. Bottom: The SFHs that are derived from
  six other independent methods and samples in the unit of number of stars formed
  per arbitrary time and spatial unit.}
  \label{fig:comparison}
\end{figure}

%%%%%%%%%%%%%%%%%%%%%%%%%%%%%%%%%%%%%%%%%%%%%%%%%%%%%%%%%%%%%%%%%%%%%%%%%%%%%%%%
\citet{2024ApJ...976...87N} derived accurate ages and metallicities for 
$\sim$$400,000$ subgiant branch stars in the solar neighbourhood using $Gaia$
and spectroscopy, building a catalogue of stellar ages that is self-consistent
across the age–metallicity space. From this age catalogue, they decompose the
SFH of the solar neighbourhood into contributions in ([Fe/H], age) space,
effectively resolving multiple epochs of star formation corresponding to
distinct metallicity regimes. Their SFH plotted in Figure~\ref{fig:comparison}
is marginalized over the whole range of metallicity. This shows great
similarity to \citet{2022Natur.603..599X}, but neither shows the most recent
star formation.

%%%%%%%%%%%%%%%%%%%%%%%%%%%%%%%%%%%%%%%%%%%%%%%%%%%%%%%%%%%%%%%%%%%%%%%%%%%%%%%%
\section{Conclusions and Future Work}
\label{sec:conclusion}
By properly propagating the uncertainties in the systematics and measurements,
this work has shown that it is possible to recover the SFH from the WDLF.
Using the uncertainties in bolometric magnitude arisen
from photometric fitting and the various assumptions made in the GCNS WDLF
catalogue, we have chosen a set of varying bin sizes such that it matches the
Nyquist sampling rate where two sampling points are required for every FWHM of
the signal. In this case, the signal is the shape of the WDLF, which ultimately
comes from the SFH. 
The modelling uncertainty is investigated through bootstrapping
the input parameters 1000 times in constructing the pWDLFs.
From the comparisons with other works on the SFH of the solar 
neighbourhood, it is obvious that our new method can retrieve non-parametric SFH.
This is achieved by carefully treating the uncertainties from noisy data using
well-established mathematical tools. The recovered SFHs agree particularly well
at an intermediate age of 0.1-9\,Gyr. We outline the most imminent issues to be
addressed as follows:

%%%%%%%%%%%%%%%%%%%%%%%%%%%%%%%%%%%%%%%%%%%%%%%%%%%%%%%%%%%%%%%%%%%%%%%%%%%%%%%%
\subsection*{Known issues}
\begin{enumerate}
    \item The GCNS WDLF has made several simplifying assumptions
    about the constituent WDs, e.g.~all are processed assuming H atmospheres and single
    star evolution. Contamination appears to be significant, and the sample is also 
    uncorrected for the complex $Gaia$ selection function \citep{2021AJ....162..142R, 2023A&A...669A..55C}, leading to poorly quantified incompleteness.
    \item The properties of the Galaxy are assumed to be static, while in reality
    the scaleheights and metallicity of the disks and stellar halo have been
    evolving through cosmic history~\citep[e.g. in][]{2025MNRAS.538.2548R}.
    There is bias in the bright end of the WDLF leading to the
    underestimation of the density. Hot WDs can be seen from the largest
    distance, so they should be the most complete in a 100\,pc sample. However,
    the single scaleheight of 365\,pc used in the entire sample of GCNS WDs
    would lead to a bias. Particularly, as pointed out in C23, the scaleheight
    of the youngest WDs could be as low as 75\,pc.
    \item There are unaccounted uncertainties from the MS progenitors,
    including their metallicity evolution and the
    binary/multiplicity fraction. They can lead to biases in the derived WDLF.
    As noted by \citet{2019ApJ...878L..11I}, the inversion
    method is sensitive to the adopted metallicity and IMF models but not the
    DA/non-DA ratio and among other choices of models. Although the pWDLF
    method does not suffer similar convergence and regularisation issues,
    this method is still sensitive to the choice of MS
    models, as would any method that requires the progenitor MS lifetimes
    in the analysis.
    \item There are uncertainties from WD modelling~\citep{2022ApJ...934...36B, 2024MNRAS.527.8687O} that are not considered,
    for example, the effects from the choice of cooling models, atmosphere
    models, and synthetic photometric models. Statistical treatment for the
    bias due to unresolved binaries containing WDs should also be taken into
    account.
    \item In the era of \textit{Gaia} astrometry, the sample size of WDs is large
    enough that we should investigate the directional dependency of the WDLFs.
    The CFHS survey~\citep{2019ApJ...887..148F} and HSC
    survey~\citep{2024MNRAS.535.3611Q} seem to reveal a WDLF dissimilar to the
    WDLFs from SDSS, SuperCOSMOS and \textit{Gaia}. The distance dependency
    should also be investigated, since the large distances probed by \textit{Gaia}
    (for the relatively bright WDs) means the line-of-sight can go through multiple
    stellar populations.
\end{enumerate}

%%%%%%%%%%%%%%%%%%%%%%%%%%%%%%%%%%%%%%%%%%%%%%%%%%%%%%%%%%%%%%%%%%%%%%%%%%%%%%%%
%When comparing with the T14 sample, we realized that there is a striking 
%difference in the SFH from the 20\,pc and 100\,pc sample at the most recent
%time.
Three important points should be investigated in the future.
The issues with the GCNS discussed earlier imply that the SFH derived in this
work is preliminary at best, and thorough selection with more recent
catalogues~\citep[e.g.][]{2021MNRAS.508.3877G} should be used in conjunction
with proper selection bias correction to maximize the potential of this method.
Secondly, arbitrary selection of SFHs reported from different works cannot draw
meaningful comparison, the by-eye comparison in the previous section should only be used
for building further hypotheses in studying SFH of the solar neighbourhood.
Corrections have to be applied differently to each of the methods to debias the
results for valid comparison, particularly for comparing SFH derived from
different stellar populations. Thirdly,
if the SFH is a strong function of distance from the Sun, then the directional
dependency should also be strong given that the Sun is not located in the middle
of the Galactic plane. Furthermore, the Sun is not located in the middle of a
major arm region of the Galaxy. The analysis should be more complex than simply
dividing the sample into zones of Galactic latitudes because, for example,
looking into the length of the Orion-Cygnus arm, where the Sun resides, should
be different to looking into the inter-arm regions; while the Perseus arm and
the Carina-Sagittarius arms are too far away for WD studies, looking into their
directions only gives us the WDs in the foreground. In the recent work by
\citet{2024MNRAS.535.3611Q}, they have found a much fainter truncation
magnitude in the Galactic WDLF using one of the deepest proper-motion and
photometry-selected samples to date. Their sample covers a sky area of
165\,deg$^2$ down to $i=24$\,mag. This may be an indication of incompleteness
in all previous surveys that have not probed sufficiently deep -- methods for
completeness correction are not useful when there are no detections at all.
Alternatively, it could be due to their assumption that $\sim90$\% of the
faintest WDs are members of the thin disc with an exponential scale height of
250\,pc, a figure too low for such an old population of stars. This will lead
to an underestimation of the generalized survey volume and an overestimation
in the spatial density, increased with respect to other surveys that have made
similar assumptions due to their significantly increased survey depth.

As pointed out in \citet{2019A&A...624L...1M} and \citet{2019ApJ...878L..11I},
the decreasing SFR trend from $\sim$10 to 6\,Gyr is consistent with the
onset of quenching observed in a cosmological context at a redshift of
z$\sim$1.8~\citep[corresponds to a lookback time of
$\sim$10\,Gyr, e.g.][]{2016MNRAS.461.1100R, 2017MNRAS.471.4155K}. It is
also compatible with the evidence of the quenching of the Milky
Way~\citep{2016A&A...589A..66H}. This is in line with the thick-disc formation
scenario attributed to a major merger event at a lookback time of 
$\sim$10\,Gyr~\citep{2018Natur.563...85H}.

Beyond systematics and directional dependence, one obvious next step is to use
the colour information. All the derivation of SFH from WDLFs of the solar 
neighbourhood were reported in single photometric passband.
Thus, all works have essentially marginalized over the colour-space. To address the
problem of degeneracy and correlated signal in the solution, simultaneously
fitting multiple WDLFs in various filters may relieve some issues with degeneracy
as the pWDLFs evolve at different rates in different wavelengths.

%%%%%%%%%%%%%%%%%%%%%%%%%%%%%%%%%%%%%%%%%%%%%%%%%%%%%%%%%%%%%%%%%%%%%%%%%%%%%%%%
\section*{Acknowledgements}
Between 2020 and 2023, MCL was supported by a European Research Council (ERC)
grant under the European Union’s Horizon 2020 research and innovation program
(grant agreement number 833031).

MCL thanks Prof. Iair Arcavi for the computing power that enabled this work.
MCL also thanks Prof. Dan Maoz for useful comments on Galactic star formation
history and Q-branch white dwarf.

This work has made use of data from the European Space Agency (ESA) mission
\textit{Gaia} (\url{https://www.cosmos.esa.int/gaia}), processed by the \textit{Gaia}
Data Processing and Analysis Consortium (DPAC,
\url{https://www.cosmos.esa.int/web/gaia/dpac/consortium}). Funding for the DPAC
has been provided by national institutions, in particular the institutions
participating in the \textit{Gaia} Multilateral Agreement.

%%%%%%%%%%%%%%%%%%%%%%%%%%%%%%%%%%%%%%%%%%%%%%%%%%%%%%%%%%%%%%%%%%%%%%%%%%%%%%%%
\section*{Data Availability}
The source code underlying this article, as well as all the data sufficient to
reproduce all of the figures in this article are available on GitHub, at 
\url{https://github.com/cylammarco/SFH-WDLF-article}. Three tables are available
in the appendix for the magnitude
resolution~(Fig.~\ref{fig:magnitude_resolution}), the WDLFs sampled at those
magnitude bins~(Fig.~\ref{fig:sfh_optimal}), and the SFHs~(Fig.~\ref{fig:sfh_optimal} \& \ref{fig:bootstrap_mean}).

%%%%%%%%%%%%%%%%%%%%%%%%%%%%%%%%%% REFERENCES %%%%%%%%%%%%%%%%%%%%%%%%%%%%%%%%%%


\bibliographystyle{mnras}
\bibliography{sfh_wd} % if your bibtex file is called example.bib

%\begin{thebibliography}{99}
%\end{thebibliography}

%%%%%%%%%%%%%%%%%%%%%%%%%%%%%%%%%%%%%%%%%%%%%%%%%%%%%%%%%%%%%%%%%%%%%%%%%%%%%%%%

%%%%%%%%%%%%%%%%%%%%%%%%%%%%%%%%%% APPENDICES %%%%%%%%%%%%%%%%%%%%%%%%%%%%%%%%%%

\appendix

%%%%%%%%%%%%%%%%%%%%%%%%%%%%%%%%%%%%%%%%%%%%%%%%%%%%%%%%%%%%%%%%%%%%%%%%%%%%%%%%
\section{Tabulated data for the bolometric magnitude resolution}
This appendix lists the bin centres and bin sizes used in the WDLFs in this work,
as found in Fig.~\ref{fig:magnitude_resolution}.

\label{appexdix:magnitude-resolution}
\begin{table}
    \centering
    \begin{tabular}{c|c}
        Magnitude [mag] & Magnitude resolution [mag] \\\hline\hline
        6.23130  & 1.3475 \\
        7.37732  & 0.9446 \\
        8.21567  & 0.7321 \\
        8.93848  & 0.7135 \\
        9.63691  & 0.6834 \\\hline
        10.30486 & 0.6525 \\
        10.93644 & 0.6107 \\
        11.50489 & 0.5262 \\
        11.96400 & 0.3920 \\
        12.30640 & 0.2928 \\\hline
        12.56487 & 0.2241 \\
        12.76920 & 0.1846 \\
        12.94318 & 0.1634 \\
        13.09824 & 0.1467 \\
        13.23833 & 0.1334 \\\hline
        13.36637 & 0.1226 \\
        13.48144 & 0.1075 \\
        13.58543 & 0.1005 \\
        13.68275 & 0.0942 \\
        13.77401 & 0.0884 \\\hline
        13.85736 & 0.0783 \\
        13.93347 & 0.0739 \\
        14.00536 & 0.0698 \\
        14.07130 & 0.0620 \\
        14.13172 & 0.0588 \\\hline
        14.18907 & 0.0558 \\
        14.24349 & 0.0530 \\
        14.29513 & 0.0503 \\
        14.34410 & 0.0477 \\
        14.39198 & 0.0481 \\\hline
        14.44142 & 0.0508 \\
        14.49437 & 0.0551 \\
        14.55188 & 0.0599 \\
        14.61699 & 0.0703 \\
        14.69506 & 0.0859 \\\hline
        14.78956 & 0.1031 \\
        14.89900 & 0.1157 \\
        15.01660 & 0.1195 \\
        15.13357 & 0.1145 \\
        15.24242 & 0.1032 \\\hline
        15.34203 & 0.0960 \\
        15.43498 & 0.0899 \\
        15.52301 & 0.0862 \\
        15.60784 & 0.0835 \\
        15.68905 & 0.0789 \\\hline
        15.76928 & 0.0815 \\
        15.85181 & 0.0835 \\
        15.93829 & 0.0894 \\
        16.03500 & 0.1040 \\
        16.14828 & 0.1226 \\\hline
        16.27756 & 0.1360 \\
        16.41573 & 0.1403 \\
        16.55448 & 0.1372 \\
        16.69153 & 0.1370 \\
        16.82786 & 0.1357 \\\hline
        16.96454 & 0.1377 \\
        17.10113 & 0.1355 \\
        17.18885 & 0.0399
    \end{tabular}
    \caption{The bin centres and bin sizes used as the magnitudes in the WDLFs
    as found in Fig.~\ref{fig:magnitude_resolution}. The bin sizes are already 
    multiplied by the factor of $2.355 /2 = 1.1775$ as explained in Section~\ref{sec:magnitude_bin_size}.}
    \label{tab:magnitude_resolution}
\end{table}

%%%%%%%%%%%%%%%%%%%%%%%%%%%%%%%%%%%%%%%%%%%%%%%%%%%%%%%%%%%%%%%%%%%%%%%%%%%%%%%%
\section{Tabulated \textit{Gaia} and reconstructed WDLF}
\label{appexdix:reconstructed-wdlf}
Table~\ref{tab:reconstructed_wdlf} provides the GCNS WDLF binned at the magnitude as tabulated in 
Appendix~\ref{appexdix:magnitude-resolution}, the best-fit WDLF and the
associated uncertainties from this work. Table~\ref{tab:sfh} best-fit SFH and bootstrapped
mean solutions and their uncertainties/standard deviations.
\begin{table*}
    \centering
    \begin{tabular}{c|cc|ccc}
         M$_{\mathrm{bol}}$ & $\Phi_{\mathrm{GCNS}}$ & $\sigma_{\Phi_{\mathrm{GCNS}}}$ & $\Phi_{\mathrm{mcmc + lsq}}$ & $-\sigma_{\Phi}$ & $+\sigma_{\Phi}$ \\
         {[mag]} & \multicolumn{2}{c|}{[N / mag / pc$^3$]} & \multicolumn{3}{c}{[N / mag / pc$^3$]} \\\hline\hline
6.231  & 2.214313e-06 & 6.553443e-07 & 3.493250e-06 & 8.881756e-07 & 9.337190e-07 \\
7.377  & 1.224601e-05 & 1.853298e-06 & 1.215019e-05 & 2.993321e-06 & 3.167152e-06 \\
8.216  & 3.384194e-05 & 3.500038e-06 & 3.585671e-05 & 8.184234e-06 & 8.473705e-06 \\
8.938  & 7.610383e-05 & 5.324397e-06 & 7.173926e-05 & 1.602056e-05 & 1.623717e-05 \\
9.637  & 1.093894e-04 & 6.500499e-06 & 1.089671e-04 & 2.394652e-05 & 2.399198e-05 \\ \hline
10.305 & 1.493797e-04 & 7.796112e-06 & 1.510974e-04 & 3.230537e-05 & 3.231323e-05 \\
10.936 & 1.795497e-04 & 8.844151e-06 & 1.787795e-04 & 4.314832e-05 & 4.257239e-05 \\
11.505 & 2.583442e-04 & 1.139297e-05 & 2.579198e-04 & 5.805083e-05 & 5.734695e-05 \\
11.964 & 3.631328e-04 & 1.569788e-05 & 3.627766e-04 & 7.589080e-05 & 7.667817e-05 \\
12.306 & 4.136290e-04 & 1.934078e-05 & 4.136644e-04 & 9.962712e-05 & 1.008819e-04 \\ \hline
12.565 & 4.869177e-04 & 2.397763e-05 & 4.873168e-04 & 1.256730e-04 & 1.258003e-04 \\
12.769 & 5.951786e-04 & 2.988469e-05 & 5.899465e-04 & 1.484138e-04 & 1.468170e-04 \\
12.943 & 6.126110e-04 & 3.152253e-05 & 6.161180e-04 & 1.679907e-04 & 1.658463e-04 \\
13.098 & 6.462091e-04 & 3.425929e-05 & 6.459097e-04 & 1.816646e-04 & 1.814984e-04 \\
13.238 & 7.232715e-04 & 3.790997e-05 & 7.104663e-04 & 1.983448e-04 & 1.990569e-04 \\ \hline
13.366 & 7.429352e-04 & 4.006353e-05 & 7.455517e-04 & 2.131595e-04 & 2.168938e-04 \\
13.481 & 8.295380e-04 & 4.526936e-05 & 8.065774e-04 & 2.335913e-04 & 2.392985e-04 \\
13.585 & 8.815936e-04 & 4.825722e-05 & 8.857653e-04 & 2.555977e-04 & 2.593295e-04 \\
13.683 & 9.957526e-04 & 5.304884e-05 & 9.737318e-04 & 2.799574e-04 & 2.812153e-04 \\
13.774 & 1.114471e-03 & 5.773182e-05 & 1.093610e-03 & 3.066491e-04 & 3.028936e-04 \\ \hline
13.857 & 1.170952e-03 & 6.298541e-05 & 1.165391e-03 & 3.139387e-04 & 3.086753e-04 \\
13.933 & 1.232247e-03 & 6.653490e-05 & 1.228966e-03 & 3.402454e-04 & 3.348548e-04 \\
14.005 & 1.336676e-03 & 7.119210e-05 & 1.328274e-03 & 4.143759e-04 & 4.086235e-04 \\
14.071 & 1.401239e-03 & 7.741532e-05 & 1.400914e-03 & 4.060630e-04 & 3.989484e-04 \\
14.132 & 1.375693e-03 & 7.868211e-05 & 1.397522e-03 & 4.116430e-04 & 4.041440e-04 \\ \hline
14.189 & 1.332457e-03 & 7.938897e-05 & 1.304960e-03 & 4.464461e-04 & 4.385935e-04 \\
14.243 & 1.365978e-03 & 8.275125e-05 & 1.336308e-03 & 4.593076e-04 & 4.446018e-04 \\
14.295 & 1.397758e-03 & 8.806223e-05 & 1.391039e-03 & 4.817621e-04 & 4.550914e-04 \\
14.344 & 1.426523e-03 & 8.948072e-05 & 1.416058e-03 & 4.867015e-04 & 4.606539e-04 \\
14.392 & 1.432661e-03 & 8.934726e-05 & 1.458203e-03 & 4.950217e-04 & 4.700236e-04 \\ \hline
14.441 & 1.538139e-03 & 9.017498e-05 & 1.496979e-03 & 4.884869e-04 & 4.591975e-04 \\
14.494 & 1.534576e-03 & 8.625141e-05 & 1.569118e-03 & 4.763298e-04 & 4.390569e-04 \\
14.552 & 1.654190e-03 & 8.587770e-05 & 1.651159e-03 & 4.479693e-04 & 3.982613e-04 \\
14.617 & 1.697711e-03 & 8.024353e-05 & 1.725936e-03 & 4.307627e-04 & 3.704841e-04 \\
14.695 & 1.829354e-03 & 7.526949e-05 & 1.843034e-03 & 4.843544e-04 & 4.146363e-04 \\ \hline
14.790 & 2.003759e-03 & 7.181462e-05 & 2.064756e-03 & 4.829561e-04 & 3.839279e-04 \\
14.899 & 1.998702e-03 & 6.776839e-05 & 2.001124e-03 & 4.139917e-04 & 3.148038e-04 \\
15.017 & 1.721019e-03 & 6.209513e-05 & 1.801798e-03 & 3.593802e-04 & 2.772322e-04 \\
15.134 & 1.459514e-03 & 5.918208e-05 & 1.477353e-03 & 3.212566e-04 & 2.624039e-04 \\
15.242 & 1.109819e-03 & 5.552692e-05 & 1.120350e-03 & 2.779693e-04 & 2.369657e-04 \\ \hline
15.342 & 8.374848e-04 & 5.274598e-05 & 8.324734e-04 & 2.365644e-04 & 2.099230e-04 \\
15.435 & 6.860083e-04 & 5.277708e-05 & 6.683195e-04 & 2.069551e-04 & 1.913429e-04 \\
15.523 & 5.938084e-04 & 5.311085e-05 & 5.657742e-04 & 1.797812e-04 & 1.713409e-04 \\
15.608 & 4.603988e-04 & 4.989976e-05 & 4.292345e-04 & 1.376085e-04 & 1.331292e-04 \\
15.689 & 2.960745e-04 & 4.312156e-05 & 2.803707e-04 & 9.071370e-05 & 9.058647e-05 \\ \hline
15.769 & 2.291564e-04 & 3.941256e-05 & 2.077402e-04 & 6.704485e-05 & 6.886399e-05 \\
15.852 & 1.435831e-04 & 3.240365e-05 & 1.511713e-04 & 4.793996e-05 & 5.137534e-05 \\
15.938 & 8.911929e-05 & 2.619568e-05 & 1.088798e-04 & 3.323241e-05 & 3.879538e-05 \\
16.035 & 8.184694e-05 & 2.548277e-05 & 7.918843e-05 & 2.269243e-05 & 3.098366e-05 \\
16.148 & 4.366694e-05 & 1.839733e-05 & 5.756888e-05 & 1.515116e-05 & 2.522133e-05 \\
16.278 & 3.719678e-05 & 1.697153e-05 & 4.278525e-05 & 1.036012e-05 & 2.125270e-05 \\ \hline
16.416 & 1.585069e-05 & 1.214068e-05 & 3.278391e-05 & 7.670646e-06 & 1.775504e-05 \\
16.554 & 2.618350e-05 & 1.846156e-05 & 2.844501e-05 & 7.217086e-06 & 1.541954e-05 \\
16.692 & 3.376085e-05 & 2.101759e-05 & 2.557141e-05 & 7.203275e-06 & 1.341742e-05 \\
16.828 & 2.925551e-05 & 2.093151e-05 & 2.352451e-05 & 7.252883e-06 & 1.173577e-05 \\
16.965 & 3.611896e-05 & 2.790730e-05 & 2.271648e-05 & 7.498655e-06 & 1.086291e-05 \\ \hline
17.101 & 1.713443e-05 & 1.939102e-05 & 2.162108e-05 & 7.439431e-06 & 1.039278e-05 \\
17.189 & 2.731905e-05 & 4.685517e-05 & 2.039458e-05 & 7.169304e-06 & 9.750354e-06
    \end{tabular}
    \caption{The GCNS WDLFs and the reconstructed WDLFs in
    Fig.~\ref{fig:sfh_optimal}.}
    \label{tab:reconstructed_wdlf}
\end{table*}

\begin{table*}
    \centering
    \begin{tabular}{c|cc|cc}
    Lookback time & $\psi_{\mathrm{best fit}}$ & $\sigma_{\psi_{\mathrm{best fit}}}$ & $\psi_{\mathrm{bootstrap}}$ & $\sigma_{\mathrm{bootstrap}}$ \\
         {[Gyr]} & \multicolumn{2}{c|}{[N / Gyr / pc$^3$]} & \multicolumn{2}{c}{[N / Gyr / pc$^3$]} \\\hline\hline
    0.069  & 1.122134e-03 & 3.744307e-07 & 1.294433e-03 & 1.826132e-04 \\
    0.110  & 1.427041e-03 & 1.905011e-07 & 1.215284e-03 & 1.567819e-04 \\
    0.163  & 1.153318e-03 & 2.113909e-07 & 1.172363e-03 & 1.027793e-04 \\
    0.238  & 3.387851e-03 & 1.838886e-07 & 2.689853e-03 & 3.816396e-04 \\
    0.338  & 3.771749e-03 & 3.244526e-07 & 3.850240e-03 & 1.725799e-04 \\ \hline
    0.470  & 4.561012e-03 & 4.110958e-07 & 3.835240e-03 & 3.635428e-04 \\
    0.645  & 1.592617e-03 & 7.777672e-07 & 2.393786e-03 & 2.950198e-04 \\
    0.865  & 1.742727e-03 & 9.624692e-07 & 1.179840e-03 & 1.805645e-04 \\
    1.100  & 4.097217e-03 & 9.730097e-07 & 3.036669e-03 & 8.218007e-04 \\
    1.320  & 2.316139e-03 & 1.115177e-06 & 3.545147e-03 & 5.418282e-04 \\ \hline
    1.520  & 2.043184e-03 & 1.091103e-06 & 2.010441e-03 & 4.119206e-04 \\
    1.705  & 4.912041e-03 & 1.228436e-06 & 3.540460e-03 & 1.033061e-03 \\
    1.885  & 2.385442e-03 & 1.307459e-06 & 3.249453e-03 & 6.645112e-04 \\
    2.065  & 2.493431e-03 & 1.509814e-06 & 2.601091e-03 & 4.225967e-04 \\
    2.245  & 3.287606e-03 & 1.667853e-06 & 2.919798e-03 & 4.524523e-04 \\ \hline
    2.425  & 2.129701e-03 & 2.130582e-06 & 2.550760e-03 & 3.525198e-04 \\
    2.600  & 2.678635e-03 & 2.132591e-06 & 2.459336e-03 & 2.940070e-04 \\
    2.770  & 2.047129e-03 & 2.152658e-06 & 2.386875e-03 & 3.603279e-04 \\
    2.940  & 3.091470e-03 & 2.280724e-06 & 2.810467e-03 & 5.637071e-04 \\
    3.110  & 3.188521e-03 & 2.189701e-06 & 3.038192e-03 & 5.351526e-04 \\ \hline
    3.275  & 2.847573e-03 & 2.189368e-06 & 3.169823e-03 & 5.636200e-04 \\
    3.435  & 3.040816e-03 & 2.233543e-06 & 3.281591e-03 & 6.247477e-04 \\
    3.595  & 5.139957e-03 & 2.023443e-06 & 3.856123e-03 & 6.555988e-04 \\
    3.750  & 3.768767e-03 & 2.271291e-06 & 3.705245e-03 & 5.505121e-04 \\
    3.900  & 2.938154e-03 & 2.611350e-06 & 3.314548e-03 & 4.598502e-04 \\ \hline
    4.050  & 3.124687e-03 & 2.543713e-06 & 3.247666e-03 & 4.347274e-04 \\
    4.200  & 2.993464e-03 & 2.681085e-06 & 3.029156e-03 & 4.811093e-04 \\
    4.350  & 2.857119e-03 & 2.481029e-06 & 2.834704e-03 & 5.299412e-04 \\
    4.500  & 2.358313e-03 & 2.792455e-06 & 2.478419e-03 & 5.002289e-04 \\
    4.655  & 2.538661e-03 & 2.012508e-06 & 2.264435e-03 & 4.385519e-04 \\ \hline
    4.825  & 1.768101e-03 & 2.434172e-06 & 1.895575e-03 & 4.544558e-04 \\
    5.020  & 2.217269e-03 & 1.961544e-06 & 1.637634e-03 & 4.494056e-04 \\
    5.250  & 1.124325e-03 & 2.730378e-06 & 1.272035e-03 & 4.451935e-04 \\
    5.540  & 6.912687e-04 & 2.944575e-06 & 8.337912e-04 & 3.629380e-04 \\
    5.945  & 7.073682e-04 & 3.508721e-06 & 5.014056e-04 & 2.000313e-04 \\ \hline
    6.560  & 5.248593e-04 & 4.222706e-06 & 4.710739e-04 & 9.332330e-05 \\
    7.445  & 1.559722e-03 & 4.009267e-06 & 1.038516e-03 & 4.259240e-04 \\
    8.365  & 1.700458e-03 & 2.029021e-06 & 2.100868e-03 & 3.335377e-04 \\
    9.050  & 1.633936e-03 & 2.677658e-06 & 1.706703e-03 & 3.779938e-04 \\
    9.520  & 1.120149e-03 & 3.471713e-06 & 1.469141e-03 & 5.284856e-04 \\ \hline
    9.870  & 5.986870e-04 & 3.959193e-06 & 1.087353e-03 & 4.029177e-04 \\
    10.155 & 9.912654e-04 & 3.803551e-06 & 8.842998e-04 & 3.795064e-04 \\
    10.400 & 9.906801e-04 & 4.231466e-06 & 7.438534e-04 & 3.731828e-04 \\
    10.620 & 1.034134e-03 & 4.522124e-06 & 6.333329e-04 & 3.871658e-04 \\
    10.820 & 4.216777e-04 & 3.416725e-06 & 5.315045e-04 & 3.878134e-04 \\ \hline
    11.010 & 2.413372e-04 & 3.554749e-06 & 4.188267e-04 & 3.649581e-04 \\
    11.200 & 1.135038e-04 & 1.762412e-06 & 3.185873e-04 & 3.276277e-04 \\
    11.395 & 9.601027e-05 & 9.326654e-07 & 2.158567e-04 & 2.525222e-04 \\
    11.610 & 4.476925e-05 & 1.152229e-06 & 1.604711e-04 & 2.364910e-04 \\
    11.860 & 2.874445e-05 & 2.686066e-06 & 1.016702e-04 & 1.898061e-04 \\ \hline
    12.145 & 1.011602e-05 & 5.461379e-06 & 5.774285e-05 & 1.335617e-04 \\
    12.450 & 9.736544e-16 & 6.354193e-06 & 3.380108e-05 & 8.350891e-05 \\
    12.755 & 4.206459e-06 & 5.614209e-06 & 2.692935e-05 & 4.518717e-05 \\
    13.050 & 3.358672e-05 & 1.038886e-06 & 3.060165e-05 & 2.912667e-05 \\
    13.330 & 6.188791e-05 & 2.605815e-07 & 4.195291e-05 & 2.631708e-05 \\ \hline
    13.590 & 8.618926e-05 & 1.141149e-07 & 5.381098e-05 & 2.363231e-05 \\
    13.825 & 8.804379e-05 & 3.782637e-07 & 8.034392e-05 & 2.864236e-05 \\
    13.965 & 4.432244e-04 & 2.288005e-08 & 3.913583e-04 & 3.212227e-05
    \end{tabular}
    \caption{The best-fit and bootstrapped mean SFH from the lower panel of
    Fig.~\ref{fig:sfh_optimal} and Fig.~\ref{fig:bootstrap_mean}.}
    \label{tab:sfh}
\end{table*}

%%%%%%%%%%%%%%%%%%%%%%%%%%%%%%%%%%%%%%%%%%%%%%%%%%%%%%%%%%%%%%%%%%%%%%%%%%%%%%%%
\section{Integration precision requirement}
\label{appexdix:integration-precision}
The integration of a WDLF at a given magnitude spans a huge dynamic range in
different parameters. In this appendix, we illustrate the importance in setting
a sufficient tolerance limit for the integrator such that a continuous WDLF can
be generated. This also shows how the precision levels can be computed for 
a given analysis using theoretical WDLFs.

\begin{figure*}
    \centering
    \begin{large}
        100 Myr population
    \end{large}
    \includegraphics[width=0.95\linewidth]{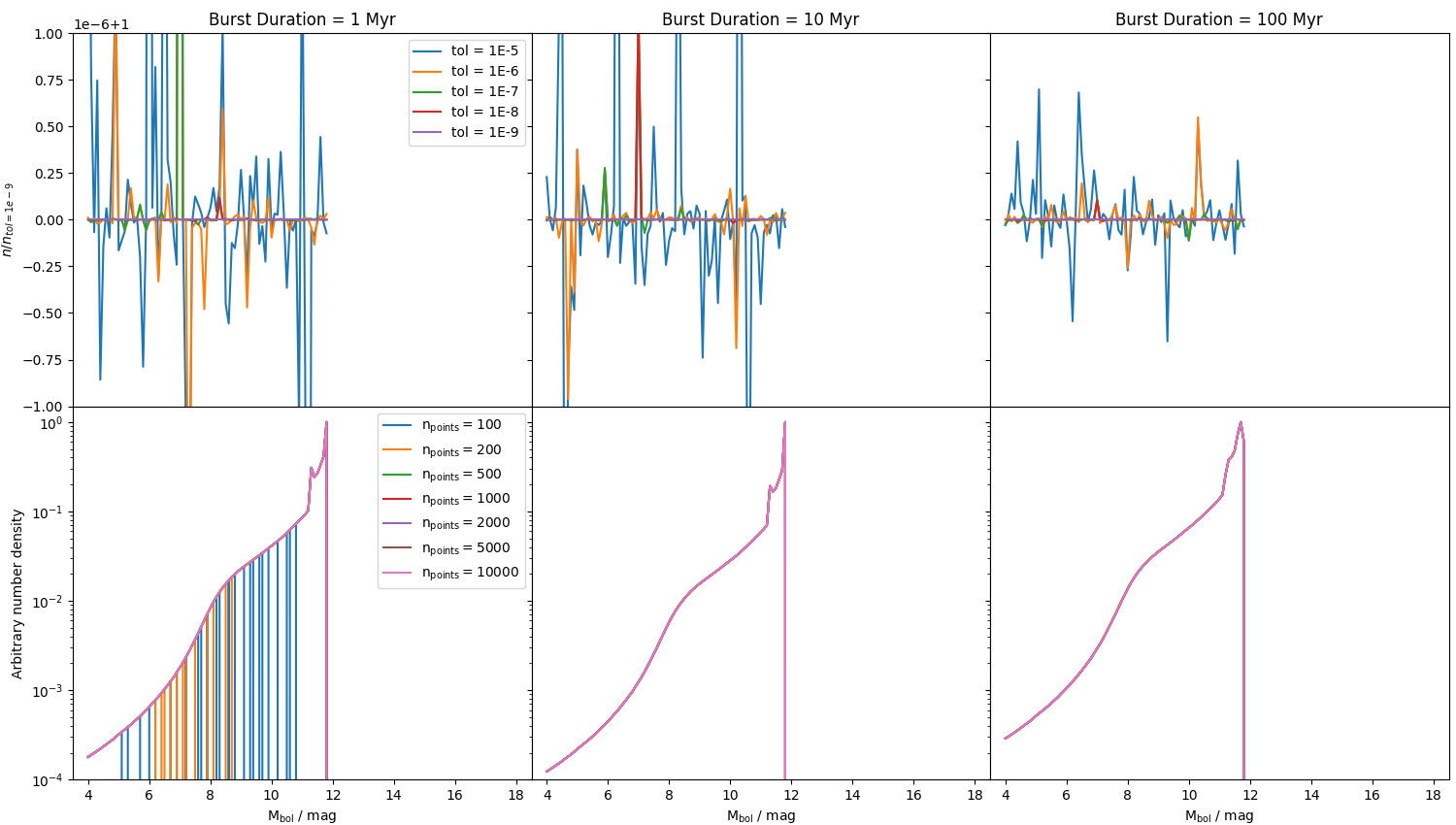} \\
    \begin{large}
        1 Gyr population 
    \end{large}
    \includegraphics[width=0.95\linewidth]{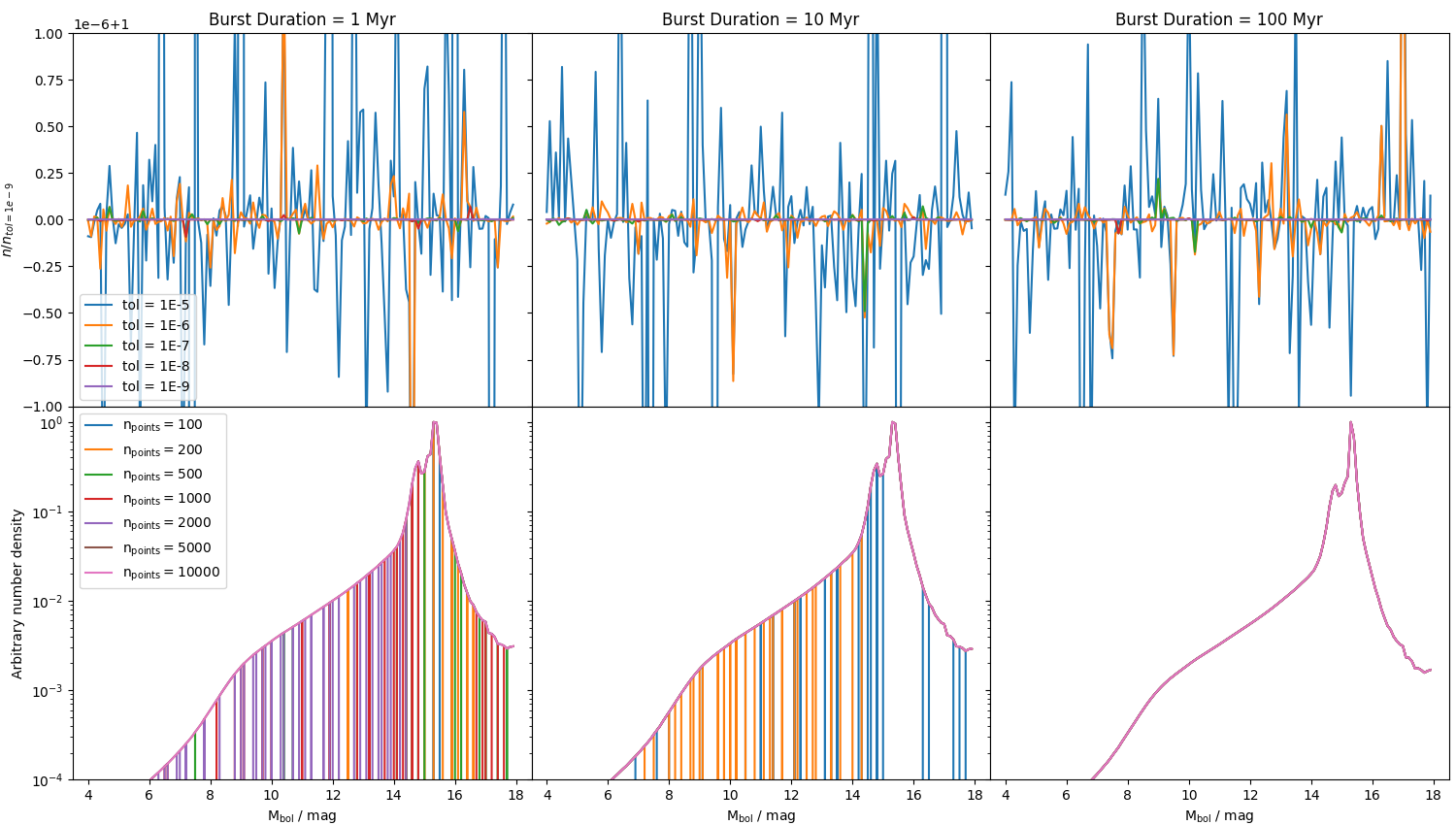}
    \caption{Top: The upper row shows the fractional difference between each of
    the 100\,Myr WDLFs integrated using \texttt{scipy.integrate.quad} with a
    relative tolerance of $10^{-5}$~(blue), $10^{-6}$~(orange),
    $10^{-7}$~(green), $10^{-8}$~(red) and $10^{-9}$~(purple) against the WDLF
    integrated using a relative tolerance of $10^{-9}$, combined with the use
    of n$_{\mathrm{points}}=10000$. They are repeated using three different
    star formation duration, at 1 Myr (left), 10 Myr (middle) and 100 Myr (right).
    The fractional differences are typically smaller than $10^{-7}$ by using a
    relative tolerance of $10^{-8}$. The lower row shows the 100 Myr WDLFs
    integrated using a fixed relative tolerance of $10^{-9}$ and
    n$_{\mathrm{points}} = 100, 200, 500, 1000, 5000$ and $10000$. It is clear
    that a n$_{\mathrm{points}}$ of at least $500$ is required to compute a
    WDLF properly. Bottom: Same as above for a 1 Gyr population. The main
    difference is that for a 1 Myr burst ($0.1\%$ of the age of the population)
    an n$_{\mathrm{points}}=5000$ is required to compute a smooth WDLF. Thus,
    for the partial WDLF we pre-compute in this work, we use a fixed
    n$_{\mathrm{points}}$ of 10000 and a relative tolerance of $10^{-10}$.}
    \label{fig:integration-precision}
\end{figure*}

% Don't change these lines
\bsp	% typesetting comment
\label{lastpage}
\end{document}